\documentclass[11pt]{article}
\usepackage[margin=0.9in]{geometry}
\usepackage[T1]{fontenc}
\usepackage{lmodern}
\usepackage{amsmath,amssymb}
\usepackage{booktabs}
\usepackage{graphicx}
\usepackage{multirow}
\usepackage{xcolor}
\usepackage[numbers,sort&compress]{natbib}
\usepackage[colorlinks=true,linkcolor=blue!50!black,citecolor=blue!50!black,urlcolor=blue!50!black]{hyperref}
\usepackage{caption}
\usepackage{float}
\newcommand{\sig}{$^{*}$}
\newcommand{\KL}{\mathrm{KL}}

\title{When Does Blockwise Additivity Break?\\
\large Diagnosing and Repairing the Additive Assumption in\\ Distillation-Based Architecture Search}
\author{Yuefeng Xu\qquad Jun Yu\\
Niigata University, Niigata, Japan\\
\texttt{f25l005c@mail.cc.niigata-u.ac.jp}\qquad \texttt{Yujun@ie.niigata-u.ac.jp}}
\date{}

\begin{document}
\maketitle

\begin{abstract}
Blockwise distillation methods for architecture search, such as DNA, DONNA, LANA and Puzzle, score each candidate
replacement block in isolation and combine the scores additively. We ask at which budgets this additivity assumption
fails, whether fine-tuning repairs the failure, what diagnoses it, and how it can be repaired cheaply. On DeiT-S,
DeiT-B, Swin-T and Qwen2.5-0.5B with seven latency budgets each, an additive integer program trails a pairwise
interaction-aware program after fine-tuning by at most 1.6 top-1 points on ImageNet-1k at the mildest budgets, but by
up to 17.5 points, and a factor of 1.9 in perplexity, at aggressive ones. After 10{,}000 fine-tuning steps 3.3--9.9
points remain on DeiT, whereas on Swin-T the gap shrinks to at most about 1.8 points within 500 steps. In a
learning-rate sweep with the standard 2{,}000 steps, 6.0--10.7 points remain on DeiT when each method's rate is chosen
by its evaluation-set accuracy. An intervention
that keeps the block library and only forbids more than two removals in any three adjacent layers shrinks the DeiT gap
from 5.8--17.5 to 0.7--5.2 points, mostly by improving the additive solution: adjacent removals are a major failure
pattern. Splitting the additivity error into a metric and a functional part, relative to additive logit changes,
indicates where a cheap logit-sum predictor can be trusted. A hybrid search that ranks moves with this predictor and
measures only the best needs no pair measurements and, after Holm adjustment, is significantly better than the pairwise
program in 10 of 14 ImageNet-1k settings, with no statistically significant losses. An ablation shows that its ranking helps except at the most
aggressive DeiT-B budget, where the predictor misleads and random ranking or ten-times-costlier evolution does better.
\par\medskip\noindent\textbf{Keywords:} Neural architecture search, model compression, blockwise distillation, vision transformers, large language models, integer programming
\end{abstract}

\section{Introduction}
\label{sec:intro}
Deploying large pretrained networks under latency constraints has motivated structured compression that removes or
replaces parts of the network~\cite{yang2023nvit,muralidharan2024minitron}.
Blockwise distillation is an attractive recipe for turning a large pretrained network into a faster one. Each block of
the parent is given a set of cheaper alternatives, for example fewer attention heads, a pruned or low-rank MLP, or no
block at all. Every alternative is trained locally to mimic the parent block's output, in the spirit of knowledge
distillation~\cite{hinton2015distilling}, and is scored once in isolation.
A child network is then assembled from these block-level scores. LANA~\cite{molchanov2022lana} and, at the scale of
large language models, Puzzle~\cite{bercovich2024puzzle} choose one alternative per layer so that a summed per-block
score is minimal under a latency or memory budget; the resulting integer program solves in seconds.
DNA~\cite{li2020dna} ranks candidate architectures by the sum of their block-wise distillation losses, and
DONNA~\cite{moons2021donna} fits a linear accuracy predictor over block-level scores and searches it with an
evolutionary algorithm. In all four, block scores enter the selection additively.

The price of this speed is an \emph{additivity assumption}: the damage done by a set of replacements is taken to be the
sum of the damage of each replacement alone. The assumption is plausible when few blocks change. It is far less
obvious when half of the network is replaced, which is exactly the regime of aggressive latency budgets. If additivity
fails there, an additive search can pick combinations that look cheap one block at a time but interact badly. The quality of the result then depends on fine-tuning to repair the damage.

In this paper we examine this assumption systematically and propose cheap ways to repair it. We compare an additive
integer linear program (ILP) with a pairwise integer quadratic program (IQP) that adds measured interaction terms, on
three vision transformers and one language model at seven latency budgets each, and ask at which budgets the additive
choice costs accuracy, whether fine-tuning recovers it, where the additivity error comes from, and how it can be
avoided without measuring all block pairs. Unlike recent studies of non-additive layer removal in language models~\cite{su2026repair,kim2026rethinking,jansen2026cbo},
which remove whole layers, we study the full blockwise distillation setting with trained replacement blocks. Our
main contributions are as follows.
\begin{enumerate}
\item We show at which compression levels the additivity assumption fails, identify the stacking of removals over
  adjacent layers as a major failure pattern, and show that whether fine-tuning repairs the damage depends on the architecture.
\item We propose a decomposition of the additivity error into a metric and a functional part that indicates when a
  cheap logit-sum predictor can replace pair measurements.
\item We propose a hybrid search that ranks candidates with this predictor and measures only the best ones; it needs no
  pair measurements and matches or outperforms the pairwise program at a fraction of its cost.
\item We propose run-length constraints, a simple linear addition to the additive program that removes most of its
  failure without any extra measurement.
\end{enumerate}
Section~\ref{sec:related} reviews related work, Section~\ref{sec:method} presents the search problem and the proposed
methods, Section~\ref{sec:exp} reports the experiments, Section~\ref{sec:analysis} analyzes the results, and
Section~\ref{sec:conclusion} concludes.

\section{Related work}
\label{sec:related}
\paragraph{Blockwise distillation search.}
DNA~\cite{li2020dna} trains supernet blocks with block-wise knowledge distillation. DONNA~\cite{moons2021donna} builds a
block library by blockwise distillation, fits a linear accuracy predictor over block-level scores and searches it
with an evolutionary algorithm rather than an integer program.
LANA~\cite{molchanov2022lana} trains alternative operations by layer-wise feature-map distillation and selects them with
an integer linear program under a latency constraint. Puzzle~\cite{bercovich2024puzzle} applies a related pipeline to
large language models with a mixed-integer program and global distillation. All of these rely, explicitly or through a
linear predictor, on per-block scores that combine additively. Our study tests this assumption directly.

\paragraph{Structured compression of transformers.}
Many attention heads can be removed with little loss~\cite{michel2019heads}, and residual networks tolerate the
removal of individual blocks~\cite{veit2016residual}. For vision transformers~\cite{dosovitskiy2021vit}, NViT
redistributes width across blocks with Hessian-based saliency~\cite{yang2023nvit}. For language models, whole layers
are removed by input--output similarity~\cite{men2025shortgpt,gromov2025deeper}, and Minitron combines pruning with
distillation~\cite{muralidharan2024minitron}. These methods also rank layers or units by individual importance scores,
so the question of how such scores combine applies to them as well.

\paragraph{Non-additive layer removal.}
Recent work on depth pruning has found that the effects of removing several layers do not add up. EvoPress observes
that per-layer importance heuristics rest on assumptions that do not hold for language models and searches
compression profiles by evolution~\cite{sieberling2025evopress}. Kim et al.\ show that multi-layer removal is
non-additive but find that subset-level search does not consistently beat one-shot pruning, whereas the calibration
objective matters more~\cite{kim2026rethinking}. Su et al.\ enumerate all layer-group subsets of four language-model
families and find positive pairwise terms that decay with distance in depth; one-shot scores then pick nearly
contiguous layers far from the optimum, and with a training-free linear repair the optimum becomes a single
contiguous block~\cite{su2026repair}. CBO formulates block removal as a constrained binary optimization with pairwise
couplings between blocks, close to our pairwise IQP, and gains most at deep compression~\cite{jansen2026cbo}. For a
vision transformer, Jung and Keuper evaluate all 4{,}095 block subsets of DINOv3 ViT-S/16 after fine-tuning and find
that keeping the first blocks is worse than an average subset~\cite{jung2026blocks}. These studies remove whole
layers. We study blockwise distillation search with a library of trained replacement blocks, on vision transformers
and a language model, decompose the additivity error, test whether fine-tuning repairs it across architectures, and
replace pair measurements by a ranked hybrid search.

\paragraph{Interaction-aware compression.}
Optimal Brain Damage scores weights with a diagonal Hessian, that is, additively~\cite{lecun1990obd}, whereas Optimal
Brain Surgeon keeps the off-diagonal terms that couple them~\cite{hassibi1993obs}. In mixed-precision quantization,
bit-widths are assigned from per-layer Hessian sensitivities~\cite{dong2019hawq}, and HAWQ-V3 selects them with an
integer linear program over per-layer scores~\cite{yao2021hawqv3}, the same additive form as LANA.
CLADO~\cite{deng2023clado} is the closest prior work: it argues that per-layer quantization errors are not independent,
estimates pairwise cross-layer terms on a small data subset, and solves an integer quadratic program. Our pairwise IQP
baseline is the block-replacement analogue of this idea. We go further in three ways. We study \emph{when} additivity
breaks as a function of the budget. We separate the part of the cross term that is only metric curvature from true
interaction. And we show that pair measurements can be replaced by ranking plus a small number of real evaluations.

\paragraph{Predictor-based and evolutionary search.}
Performance predictors are widely used to make neural architecture search cheap~\cite{white2021predictors}. Regularized
evolution~\cite{real2019regularized} and random search~\cite{li2019randomsearch} are strong, simple baselines. We use
both on measured scores as equal-cost baselines for our hybrid search.

\section{Method}
\label{sec:method}
We first define the search problem and the two programs compared, then the diagnostic decomposition, and finally the
two proposed repairs.

\subsection{Blockwise replacement search}
\label{sec:problem}
A pretrained parent network is compressed by replacing some of its blocks with cheaper alternatives. Scores are
computed on a held-out calibration set (\emph{calib}), and a latency table gives the cost of every option at every
layer; both are specified for each model in Section~\ref{sec:setup}.

\paragraph{Block library.}
At each of the 12 layers, the search chooses between the original block and seven replacement options: half of
the attention heads (\texttt{attn\_heads\_half}), no attention (\texttt{attn\_skip}), an MLP retaining 50\% or 25\% of
its hidden units (\texttt{mlp\_prune50}, \texttt{mlp\_prune25}), a low-rank MLP with 50\% of the parameters
(\texttt{mlp\_lowrank50}), no MLP (\texttt{mlp\_skip}), and no block at all (\texttt{block\_skip}). The search space therefore has $8^{12}\approx 6.9\times10^{10}$ structures. Alternatives are initialized from the
parent's weights. Each is trained for 3{,}000 steps (batch 128, AdamW, learning rate $3\times10^{-4}$) to reproduce the
parent block's output from the parent block's input. With $x$ the token features entering the block, $f_p$ the parent
block and $f_c$ the candidate, the loss is $\|f_c(x)-f_p(x)\|^2/\max(\|f_p(x)-x\|^2,10^{-8})$ (means over the batch),
which normalizes by the residual update. \texttt{block\_skip} has no parameters and is not trained; for
\texttt{attn\_skip} and \texttt{mlp\_skip} the remaining sub-layer is trained.

\paragraph{Scores.}
The score of a structure is the mean $\KL(p_{\text{parent}}\,\|\,p_{\text{child}})$ over the 1000-way softmax on
\emph{calib}. We measure all $12\times7=84$ single replacements; 1{,}389 pairs, namely all 49 candidate pairs for the
21 layer pairs at distance $\le 2$ plus 8 sampled candidate pairs for each of the 45 farther layer pairs; and, for
analysis only, 400 random structures with 2--6 replaced blocks.

\paragraph{Search programs.}
All searches use the same scores and latency table.
\begin{itemize}
\item \emph{Additive ILP} minimizes $\sum_i s(i,c_i)$ under the budget, as in LANA and Puzzle.
\item \emph{Pairwise IQP} adds a measured interaction term $K(i,c_i,j,c_j)=s(i,c_i,j,c_j)-s(i,c_i)-s(j,c_j)$ for every
  measured pair. The products are linearized in the standard way. Pairs that were not measured (most option pairs of
  layers more than two apart, see above) enter with $K=0$, so the IQP is additive for them; it also
  ignores interactions of three or more blocks.
\item \emph{Greedy} starts from the parent and repeatedly applies the single replacement with the smallest increase in
  additive score per unit of latency saved until the budget is met. It serves only as a starting point below.
\end{itemize}
Both integer programs are solved with HiGHS~\cite{huangfu2018highs} through SciPy with a 120\,s time limit. Every solve
reported here terminated with an optimal status within HiGHS's default relative MIP gap of $10^{-4}$, after at most
16\,s.

\subsection{Decomposing the additivity error}
\label{sec:decomp}
Let $z\in\mathbb{R}^{1000}$ be the parent's logits on an image, $\mathrm{sm}$ the softmax, and $\delta_i$ the change
in logits caused by the replacement selected at layer $i$ alone (measured with the other blocks original; strictly
$\delta_{i,c_i}$). For a structure $A$ we compare three quantities:
\begin{align*}
\text{add}(A) &= \textstyle\sum_{i\in A}\KL\big(\mathrm{sm}(z)\,\|\,\mathrm{sm}(z+\delta_i)\big),\\
\text{lin}(A) &= \KL\big(\mathrm{sm}(z)\,\|\,\mathrm{sm}(z+\textstyle\sum_{i\in A}\delta_i)\big),\\
\text{act}(A) &= \KL\big(\mathrm{sm}(z)\,\|\,\mathrm{sm}(z_A)\big),
\end{align*}
each averaged over \emph{calib}. Here $z_A$ are the measured logits of the child. The additivity error splits exactly
into
\begin{equation}
\text{act}-\text{add} \;=\; \underbrace{(\text{lin}-\text{add})}_{\text{metric}} \;+\;
\underbrace{(\text{act}-\text{lin})}_{\text{functional}}.
\end{equation}
The metric part would be present even if the blocks did not interact at all, that is, if their logit changes simply
added up. It is a property of the score, not of the network. The functional part is the difference between the KL of the
actual child logits and that of the summed-logit approximation. It captures the score-level effect of non-additive
logit changes. A functional part of zero does not prove that the logits are additive, since different logits can have
the same KL. The split is a decomposition of the \emph{score} relative to one particular counterfactual, additive
logit changes; it is exact by construction but not unique, and it is not a causal decomposition of the network's
behavior. A different counterfactual (for example, additive changes of hidden features or of log-probabilities) would
split the same error differently. We use it as a diagnostic of when a predictor built on that counterfactual can work.

\paragraph{Why the metric part is often positive.}
A second-order expansion of the KL around the parent logits gives
$\KL(\mathrm{sm}(z)\,\|\,\mathrm{sm}(z+\delta))\approx\tfrac12\delta^\top F\delta$, where $p=\mathrm{sm}(z)$ and
$F=\mathrm{diag}(p)-pp^\top$ is the Fisher matrix of the softmax. With $\delta=\sum_i\delta_i$,
\begin{equation}
\text{lin}(A)-\text{add}(A)\;\approx\;\sum_{i<j}\delta_i^\top F\,\delta_j .
\end{equation}
The metric part is therefore approximately a Fisher-weighted cross term. Individual terms can be negative; the sum is
positive when the logit changes of different blocks are, on aggregate, positively aligned under $F$, for example when
several of them lower the parent's top-class logit. This is one mechanism for the super-additivity of most structures
(Table~\ref{tab:additivity}).

\paragraph{Remainder and validity.}
The expansion is only an interpretation. With $A=\log\sum_k e^{z_k}$, the KL equals $A(z+\delta)-A(z)-p^\top\delta$,
and along $g(t)=A(z+t\delta)$ the third derivative is the third central moment of the entries of $\delta$ under
$\mathrm{sm}(z+t\delta)$, so $|g'''|\le\omega\,g''$ with $\omega=\max_k\delta_k-\min_k\delta_k$. Integrating gives,
with $\phi(a)=2(e^a-1-a)/a^2$,
\begin{equation}
\phi(-\omega)\;\le\;\frac{\KL\big(\mathrm{sm}(z)\,\|\,\mathrm{sm}(z+\delta)\big)}{\tfrac12\delta^\top F\delta}\;\le\;\phi(\omega).
\end{equation}
The relative error of the quadratic term is therefore of order $\omega/3$: it is small when every logit change is well
below one nat ($\phi(\pm0.3)\in[0.90,1.11]$) and loose otherwise. Section~\ref{sec:q2} measures how far this reading holds at the budgets searched.

\subsection{Logit-sum predictor and hybrid search}
\label{sec:q3}
\paragraph{Logit-sum predictor.}
The lin predictor needs only the 84 single-block runs. Unlike the additive score, it is not linear in the choices, so we
optimize it by steepest-descent local search. The moves change one layer or two layers at once within the budget, and
the search starts from the ILP and greedy solutions. We call the result \emph{lin}.

\paragraph{Hybrid search.}
The hybrid search uses lin only as a ranker and uses directly measured calibration KL to select moves.
\begin{enumerate}
\item Start from the best of \{lin, ILP, greedy\}, judged by real KL on a 1{,}000-image subset of \emph{calib} (every
  fifth image).
\item At each step, rank all feasible one- and two-layer moves by lin. Measure the real subset KL of the 12 best moves
  not yet measured.
\item Move to the best of them if it improves on the current structure, and re-rank from the new structure.
  Otherwise measure the next 12, up to four chunks, and then stop.
\end{enumerate}
Measured structures are cached and never re-measured; the evaluation count includes the starting structures. The
hybrid requires no dedicated pairwise measurements or precomputed pair table.

\subsection{Run-length constraints}
\label{sec:runlen_method}
An additive objective cannot penalize replacements that are harmless alone but harmful when stacked over adjacent
layers. We therefore propose run-length constraints, which forbid such stacks directly, keep the library unchanged and
keep the program linear. They can be added to either integer program; we add them to \emph{both}.
\begin{itemize}
\item \emph{Same option, $R$}: in every window of $R+1$ adjacent layers, at most $R$ layers carry the same replacement
  option ($\sum_{i=s}^{s+R}x_{i,o}\le R$ for every option $o$ and window start $s$).
\item \emph{Removal class, $R$}: \texttt{attn\_skip}, \texttt{mlp\_skip} and \texttt{block\_skip} are treated as one
  class, and in every window of $R+1$ adjacent layers at most $R$ layers carry any of them.
\end{itemize}
The second constraint closes a loophole of the first: under ``same option'' the additive ILP can alternate similar
removals, for example \texttt{block\_skip} on layers 1--2 and 4--5 with \texttt{mlp\_skip} on layer 3 (DeiT-B, 55\%,
$R=2$). Neither variant needs additional measurements. If long runs of removals drive the additivity failure, the gap
between the two programs should shrink when both obey the cap.

\section{Experiments}
\label{sec:exp}
\subsection{Experimental setup}
\label{sec:setup}
\paragraph{Models and data.}
The parent models are DeiT-S and DeiT-B~\cite{touvron2021deit}, data-efficient variants of ViT~\cite{dosovitskiy2021vit}, pretrained on ImageNet-1k, taken from
timm~\cite{wightman2019timm} and used without retraining. We work on ImageNet-100, formed from every 10th
ImageNet-1k class in the standard label order (labels $0, 10, \dots, 990$)~\cite{russakovsky2015imagenet}. The last
fifty training images of each class in file-name order (5{,}000 images, \emph{calib}) are held out. They are used for every score and every
search decision. The remaining training images are used for blockwise distillation and fine-tuning. Section~\ref{sec:in1k} repeats the
main experiments on the full ImageNet-1k. In the ImageNet-100 experiments we report top-1 accuracy on two sets:
\begin{itemize}
\item \emph{val100}: the 5{,}000 official validation images of the 100 classes, 100-way, obtained by restricting the
  argmax to the 100 class logits. This is the main metric.
\item \emph{val1k}: the full 50{,}000-image ImageNet-1k validation set, 1000-way. This is a secondary metric that probes
  how far a child generalizes beyond the classes it was fine-tuned on.
\end{itemize}
The parents reach 95.76\% / 79.87\% (DeiT-S) and 96.30\% / 81.97\% (DeiT-B) on val100 / val1k.

\paragraph{Latency.}
A per-block latency table was measured on an RTX 5090 (batch 64, 20 warm-up and 100 timed runs). A budget of $x\%$
means that the summed block latency of the child is at most $x\%$ of the parent's summed block latency. For 20 random
structures, the summed table has a mean absolute relative prediction error of 1.2\% (maximum 2.8\%). We also timed every
final structure end to end; there the mean absolute relative error is 0.8\% (DeiT-S; maximum 2.7\%) and 2.5\% (DeiT-B;
maximum 6.0\%). Because the patch embedding and the classifier head are not replaced, the end-to-end latency fraction is a
few points above the budget (Table~\ref{tab:latency}).

\paragraph{Fine-tuning and statistics.}
Each searched child is fine-tuned by global distillation from the parent: a KL loss on the 1000-way softmax, 2{,}000
steps, batch 128, AdamW, learning rate $5\times10^{-5}$, weight decay 0.05, 100 warm-up steps. This is done with three
seeds. Differences between two searches are reported as the seed-averaged top-1 difference with a 95\% paired bootstrap
confidence interval over images (10{,}000 resamples, per-image correctness averaged over seeds~\cite{efron1993bootstrap}).
The bootstrap resamples validation images while holding the three trained seed realizations fixed, so the intervals
quantify image-sampling uncertainty conditional on these runs; they do not capture the variability of retraining.
Each search is run once per budget; random and evolutionary searches use one fixed seed, and
Section~\ref{sec:seeds} repeats them with three seeds. In tables, \sig\ marks a difference whose
unadjusted interval excludes zero. Because the paper makes many comparisons, every count of significant results in
the text and in win/tie/loss (W/T/L) tables is Holm-adjusted~\cite{holm1979} within a family of comparisons (for
example, all IQP $-$ ILP comparisons on ImageNet-1k, or all comparisons of the hybrid with the other searches), using
two-sided paired $z$-tests over images on the seed-averaged per-image correctness; where the adjustment changes a
statement we say so. A tie denotes a difference that is not significant after adjustment; it does not establish
equivalence. Alongside the seed-averaged differences we report per-seed differences (same fine-tuning seed index)
for the new experiments; in every comparison reported in Sections~\ref{sec:runlen} and~\ref{sec:ablation}, the three
per-seed differences have the same sign as the mean whenever the mean is significant. Exact per-seed McNemar tests~\cite{mcnemar1947} of IQP $-$ ILP are significant for every seed at budgets of 55\%
and below ($p<10^{-13}$).

\paragraph{Equal-cost baselines.}
The hybrid differs from the IQP in two ways: it ranks with lin, and it measures real structures. To separate the two
effects we ran two baselines with exactly the same number of real evaluations per budget, the same subset and the same
starting points: \emph{random}, the best of the starts and feasible structures sampled uniformly (each layer's option
drawn uniformly, rejected if over budget), and \emph{evolution}, regularized evolution~\cite{real2019regularized} with
population 24, tournament size 6, and mutations of one or two layers.

\subsection{Results on ImageNet-100}
\label{sec:q1}
\begin{table*}[t]
\centering
\caption{Additivity of single-block KL scores on 400 random structures with $k=2..6$ replaced blocks (80 per $k$;
\emph{calib}). ``Super-additive'' means that the measured KL exceeds the sum of the single scores (add). MAE is the mean
absolute difference between the measured KL and add.}
\label{tab:additivity}
\small
\setlength{\tabcolsep}{4pt}
\begin{tabular}{lcccccc}
\toprule
model & Kendall $\tau$ (sum) & $\tau$ (+ pair terms) & super-additive & median act./add & MAE $k{=}2$ & MAE $k{=}6$\\
\midrule
DeiT-S & 0.894 & 0.952 & 90\% & 1.16 & 0.027 & 0.228\\
DeiT-B & 0.894 & 0.956 & 79\% & 1.09 & 0.003 & 0.246\\
\bottomrule
\end{tabular}
\end{table*}

\paragraph{Before fine-tuning.}
Table~\ref{tab:additivity} shows that the sum of single scores ranks random structures reasonably well (Kendall
$\tau=0.89$). However, the error grows quickly with the number of replaced blocks, and it is biased: 79--90\% of the
structures are worse than predicted. Adding the measured pair terms raises $\tau$ to 0.95. Interaction is concentrated
between nearby layers. On DeiT-S the mean relative pair term, $|K(i,c_i,j,c_j)|/|s(i,c_i)+s(j,c_j)|$, is 0.167 at layer
distance $\le2$ and 0.030 beyond.

\begin{figure*}[t]
\centering
\includegraphics{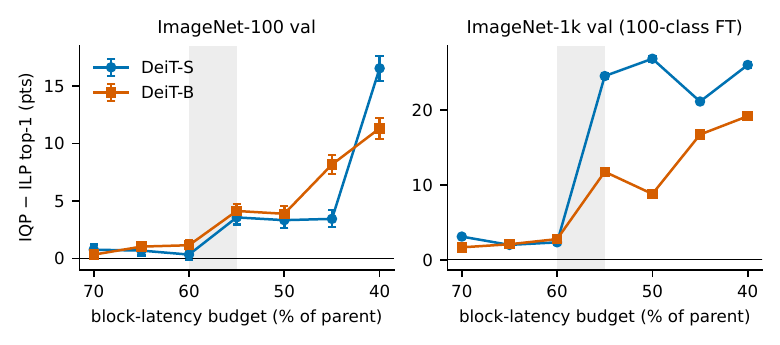}
\caption{Top-1 gain of the pairwise IQP over the additive ILP after fine-tuning (mean of three seeds, 95\% paired
bootstrap interval). The gain is about one point or less down to 60\% latency and jumps between 60\% and 55\% (shaded)
for both models.}
\label{fig:gap}
\end{figure*}

\begin{table*}[t]
\centering
\caption{IQP $-$ ILP top-1 accuracy difference after fine-tuning (percentage points; three seeds; 95\% paired
bootstrap interval; \sig: interval excludes 0). For val1k only point estimates are shown; all val1k intervals exclude 0
and are at most 0.8 points wide. The DeiT-B 70\% val100 interval has a lower bound of 0 (to numerical precision) and is
not counted as significant.}
\label{tab:gap}
\small
\begin{tabular}{lcccc}
\toprule
budget & DeiT-S val100 & DeiT-B val100 & DeiT-S val1k & DeiT-B val1k\\
\midrule
70\% & $+0.77$ [$+0.32$, $+1.21$]\sig & $+0.33$ [$0.00$, $+0.67$] & $+3.09$\sig & $+1.65$\sig\\
65\% & $+0.69$ [$+0.26$, $+1.11$]\sig & $+1.03$ [$+0.65$, $+1.43$]\sig & $+1.98$\sig & $+2.07$\sig\\
60\% & $+0.34$ [$-0.11$, $+0.81$]     & $+1.15$ [$+0.71$, $+1.59$]\sig & $+2.32$\sig & $+2.75$\sig\\
55\% & $+3.57$ [$+2.95$, $+4.21$]\sig & $+4.13$ [$+3.51$, $+4.76$]\sig & $+24.56$\sig & $+11.76$\sig\\
50\% & $+3.33$ [$+2.65$, $+4.02$]\sig & $+3.89$ [$+3.23$, $+4.54$]\sig & $+26.88$\sig & $+8.79$\sig\\
45\% & $+3.45$ [$+2.71$, $+4.19$]\sig & $+8.17$ [$+7.38$, $+8.99$]\sig & $+21.15$\sig & $+16.74$\sig\\
40\% & $+16.55$ [$+15.47$, $+17.64$]\sig & $+11.32$ [$+10.40$, $+12.25$]\sig & $+26.04$\sig & $+19.19$\sig\\
\bottomrule
\end{tabular}
\end{table*}

\paragraph{After fine-tuning.}
Figure~\ref{fig:gap} and Table~\ref{tab:gap} compare the two searches after fine-tuning. At 70\%--60\% latency the
interaction-aware search is ahead by at most 1.2 points on val100. The difference is significant in four of six
settings, but it is small. Between 60\% and 55\% the gap jumps, for both models, to 3.6--4.1 points. It grows to 11--17 points at
40\%. The val1k gap shows the same transition, from 1.6--3.1 points to 8.8--26.9 points. This is consistent with the
additive children losing much of their accuracy on the 900 classes not seen during fine-tuning.

\paragraph{Longer fine-tuning.}
Fine-tuning the DeiT-B children of the three mild budgets for 10{,}000 instead of 2{,}000 steps shrinks the gap only
slightly; it stays significant at all three budgets (Table~\ref{tab:ft10k}). Section~\ref{sec:curve}
repeats this test at the aggressive budgets.

\paragraph{Cheap repairs.}
On DeiT-S, lin substantially improves over the ILP at most budgets and its mean accuracy exceeds the IQP's at
  five of seven budgets, without dedicated pair measurements (Table~\ref{tab:methods}). It equals the ILP at 65\%,
  where both select the same structure. On DeiT-B, lin helps at mild budgets but falls 1.05, 4.32 and 5.58 points behind the IQP at 50\%, 45\% and 40\%.
  This is the regime where the functional part dominates (Sec.~\ref{sec:q2}): lin selects structures whose real KL is
  far above its prediction (Fig.~\ref{fig:calib}).

\begin{figure*}[t]
\centering
\includegraphics{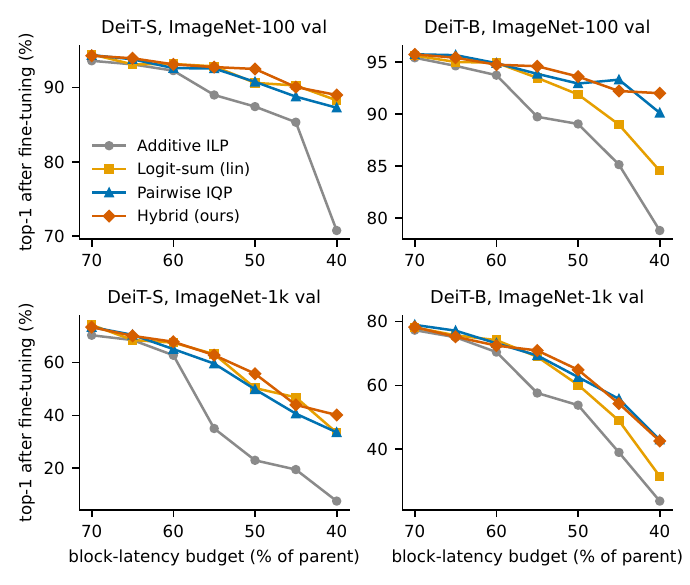}
\caption{Top-1 after fine-tuning for the four searches (three seeds). Top: val100; bottom: val1k.}
\label{fig:methods}
\end{figure*}

\begin{table}[t]
\centering
\caption{val100 top-1 after fine-tuning (three seeds) and hybrid $-$ IQP with 95\% paired bootstrap interval.}
\label{tab:methods}
\footnotesize
\setlength{\tabcolsep}{2.5pt}
\begin{tabular}{llccccc}
\toprule
model & budget & ILP & lin & IQP & hybrid & hybrid $-$ IQP\\
\midrule
\multirow{7}{*}{DeiT-S}
& 70\% & 93.59 & 94.47 & 94.36 & 94.27 & $-0.09$ [$-0.53$, $+0.34$]\\
& 65\% & 93.11 & 93.11 & 93.80 & 93.93 & $+0.13$ [$-0.24$, $+0.49$]\\
& 60\% & 92.27 & 93.16 & 92.61 & 93.11 & $+0.51$ [$+0.09$, $+0.94$]\sig\\
& 55\% & 88.99 & 92.85 & 92.56 & 92.75 & $+0.19$ [$-0.23$, $+0.61$]\\
& 50\% & 87.42 & 90.62 & 90.75 & 92.49 & $+1.74$ [$+1.19$, $+2.30$]\sig\\
& 45\% & 85.32 & 90.27 & 88.77 & 90.05 & $+1.29$ [$+0.71$, $+1.87$]\sig\\
& 40\% & 70.73 & 88.27 & 87.27 & 88.99 & $+1.71$ [$+1.13$, $+2.30$]\sig\\
\midrule
\multirow{7}{*}{DeiT-B}
& 70\% & 95.39 & 95.59 & 95.72 & 95.71 & $-0.01$ [$-0.31$, $+0.29$]\\
& 65\% & 94.61 & 94.99 & 95.64 & 95.39 & $-0.25$ [$-0.59$, $+0.09$]\\
& 60\% & 93.71 & 94.91 & 94.87 & 94.72 & $-0.15$ [$-0.55$, $+0.25$]\\
& 55\% & 89.71 & 93.45 & 93.84 & 94.57 & $+0.73$ [$+0.27$, $+1.19$]\sig\\
& 50\% & 89.02 & 91.86 & 92.91 & 93.58 & $+0.67$ [$+0.19$, $+1.16$]\sig\\
& 45\% & 85.11 & 88.96 & 93.28 & 92.19 & $-1.09$ [$-1.65$, $-0.54$]\sig\\
& 40\% & 78.77 & 84.51 & 90.09 & 91.98 & $+1.89$ [$+1.29$, $+2.48$]\sig\\
\bottomrule
\end{tabular}
\end{table}

\paragraph{Accuracy.}
After fine-tuning (Fig.~\ref{fig:methods}, Table~\ref{tab:methods}), the hybrid is significantly better than the IQP in
4 of 14 settings after Holm adjustment (7 unadjusted), not significantly different in 9, and significantly worse in one
(DeiT-B 45\%, $-1.09$). It is significantly better than the additive ILP in 12 of 14 settings. On val1k the picture is
more mixed: the hybrid is significantly ahead of the IQP in 7 settings and behind in 5, mostly the mild DeiT-B budgets.

\paragraph{Cost.}
Beyond the 84 single runs shared by all methods, the IQP needs 1{,}389 pair measurements on 5{,}000 images, which is
6.9\,M forward images. The hybrid measured 930 structures (DeiT-S) and 1{,}843 structures (DeiT-B) on 1{,}000 images
over all seven budgets, that is 13.4\% and 26.5\% of the IQP pair stage (Table~\ref{tab:cost}).
Measured on one RTX~5090 on full ImageNet-1k, the IQP pair stage takes 10.6 (DeiT-S) and 27.3 (DeiT-B) minutes,
whereas the hybrid search, including its lin ranking, takes 1.5 and 4.2 minutes over all seven budgets; the shared
single-block stage takes 0.6 and 1.7 minutes, and each integer program solves in at most 16\,s. The hybrid's cost
grows with the number of budgets, whereas the IQP's pair table is reused; at the same cost per budget, the two break
even after about 26--52 budgets.

\paragraph{Equal-cost baselines.}

Table~\ref{tab:baselines} gives these comparisons. On val100, after Holm adjustment, equal-cost evolution on
measured KL is not significantly different from the hybrid in any of the 14 settings, and random sampling in 12
(2 wins of the hybrid). On DeiT-S, random and evolution often kept the lin start, so part of their strength there is
inherited from lin. These results suggest that \emph{direct evaluation of complete structures} accounts for a
substantial part of the improvement over the IQP. Because all baselines share starting points that include lin, this
comparison does not isolate the full contribution of the logit-sum predictor; Section~\ref{sec:ablation} ablates the
start and the ranker separately.

The ranking appears to matter for generalization. On val1k the hybrid beats evolution in 11 of 14 settings. On DeiT-B the
margin is $+3$ to $+9$ points at six of seven budgets. Proposing moves that are good under the logit-sum model appears
to favor structures whose errors are small across all 1000 classes, not just on the measured subset.

\paragraph{Further studies.}
Two follow-up studies in the appendix probe what limits the searches beyond the hybrid.
A stronger search (\emph{hybrid~v2}, Appendix~\ref{sec:hybrid2}) finds structures with lower
pre-fine-tuning KL but no consistent gain after fine-tuning, and a fine-tuning-aware score
(Appendix~\ref{sec:ftproxy}) ranks the good structures no better than calibration top-1 before
fine-tuning. Among good structures, the fidelity of the score rather than the search limits further gains.

\subsection{Full ImageNet-1k}
\label{sec:in1k}
The experiments above fine-tune the children on 100 classes, so val1k could only be a secondary metric. We therefore
repeated the main pipeline on the full ImageNet-1k for both models, with no change of method or hyperparameters. The
block library is distilled on the full training set. Five training images per class (5{,}000 images) are held out as
\emph{calib}. We measure all 84 single replacements, the same set of 1{,}389 replacement pairs defined in Section~\ref{sec:setup},
and 400 random structures. All six searches (ILP,
IQP, lin, hybrid, random, evolution) are run at the seven budgets. Every child is fine-tuned on all 1000 classes with the
recipe of Section~\ref{sec:setup} and three seeds, and is evaluated 1000-way on the 50{,}000 official validation images.
This is now the only metric.

\paragraph{Additivity and its decomposition.}
{\sloppy The \emph{calib} statistics are similar to those obtained on ImageNet-100. On the 400 random structures, the plain sum ranks with Kendall
$\tau$ of 0.887 on DeiT-S and 0.886 on DeiT-B, and adding pair terms raises $\tau$ to 0.951 and 0.954. Of the
structures, 95\% and 92\% are super-additive, with median act./add ratios of 1.22 and 1.21. Interaction is again local: the mean relative pair term is 0.19 at layer
distance $\le2$ and 0.03--0.04 beyond. The KL decomposition reproduces the contrast of Table~\ref{tab:decomp}. The metric
share is 63\% (pairs) and 67\% (structures) on DeiT-S, against 36\% and 34\% on DeiT-B, so on the larger model most of
the mean error is again attributed to the functional component.\par}

\begin{table*}[t]
\centering
\caption{Full ImageNet-1k: top-1 (\%) on the 50{,}000 validation images after fine-tuning on all 1000 classes (mean of
three seeds; seed standard deviation 0.03--0.16). Gap: IQP $-$ ILP with 95\% paired bootstrap interval; every interval
excludes 0. Boldface marks the highest mean in each row; it does not indicate statistical significance. The parents
reach 79.87\% (DeiT-S) and 81.97\% (DeiT-B).}
\label{tab:in1k}
\small
\setlength{\tabcolsep}{4pt}
\begin{tabular}{llccccccc}
\toprule
model & budget & ILP & IQP & lin & random & evolution & hybrid & IQP $-$ ILP [95\% CI]\\
\midrule
\multirow{7}{*}{DeiT-S}
 & 70\% & 74.28 & 75.45 & 74.93 & 74.89 & 74.89 & \textbf{75.77} & $+1.17$ [$+0.95$, $+1.40$]\\
 & 65\% & 73.34 & 74.23 & 73.29 & 73.05 & 73.50 & \textbf{74.53} & $+0.90$ [$+0.66$, $+1.13$]\\
 & 60\% & 72.83 & 72.04 & 72.88 & 72.83 & 72.83 & \textbf{73.21} & $-0.79$ [$-1.04$, $-0.53$]\\
 & 55\% & 63.21 & 70.24 & \textbf{71.41} & \textbf{71.41} & \textbf{71.41} & 71.30 & $+7.03$ [$+6.71$, $+7.34$]\\
 & 50\% & 58.19 & 67.85 & 69.26 & 69.26 & 69.26 & \textbf{69.34} & $+9.66$ [$+9.34$, $+9.99$]\\
 & 45\% & 56.52 & 63.06 & 65.85 & 65.85 & 65.85 & \textbf{66.27} & $+6.54$ [$+6.21$, $+6.88$]\\
 & 40\% & 42.96 & 60.46 & 58.57 & 58.57 & 61.83 & \textbf{62.27} & $+17.50$ [$+17.11$, $+17.89$]\\
\midrule
\multirow{7}{*}{DeiT-B}
 & 70\% & 79.27 & 79.88 & 79.27 & 77.68 & 79.16 & \textbf{80.07} & $+0.61$ [$+0.43$, $+0.77$]\\
 & 65\% & 77.15 & 78.62 & \textbf{78.88} & \textbf{78.88} & \textbf{78.88} & \textbf{78.88} & $+1.47$ [$+1.27$, $+1.67$]\\
 & 60\% & 75.50 & 77.11 & \textbf{78.29} & 75.14 & 77.26 & 77.82 & $+1.61$ [$+1.38$, $+1.84$]\\
 & 55\% & 69.76 & 75.55 & 74.65 & 74.72 & 75.13 & \textbf{76.71} & $+5.79$ [$+5.52$, $+6.07$]\\
 & 50\% & 66.13 & \textbf{74.45} & 69.81 & 70.19 & 72.51 & 74.29 & $+8.32$ [$+8.03$, $+8.62$]\\
 & 45\% & 54.95 & 71.05 & 62.88 & 70.92 & 70.30 & \textbf{72.13} & $+16.10$ [$+15.75$, $+16.45$]\\
 & 40\% & 49.21 & 63.83 & 53.00 & 64.73 & \textbf{67.17} & 65.54 & $+14.62$ [$+14.27$, $+14.96$]\\
\bottomrule
\end{tabular}
\end{table*}

\paragraph{The transition.}
Table~\ref{tab:in1k} shows the same break as on ImageNet-100. Down to 60\% latency the IQP and the ILP differ by at most
1.6 points. On DeiT-S the ILP is even 0.8 points ahead at 60\%. Between 60\% and 55\% the gap jumps to 7.0 (DeiT-S) and
5.8 points (DeiT-B), and it reaches 6.5--17.5 points at 50--40\%. The jump is larger than on val100 of ImageNet-100
(3.6--4.1 points at 55\%); this larger gap is observed when the children are fine-tuned and evaluated on all 1000
classes.

\paragraph{Cheap repairs.}
The behavior of lin again follows the decomposition. On DeiT-S, where most of the error is metric, lin's mean accuracy is
1.2--2.8 points above the IQP's at 55\%--45\% without dedicated pair measurements. On DeiT-B, where most of the error is
functional, lin falls 4.6--10.8 points behind the IQP at 50\%--40\%. The hybrid removes this weakness
(Table~\ref{tab:in1k_wtl} gives all Holm-adjusted counts). After Holm adjustment it significantly outperforms the IQP in 10 of 14 settings and is
never significantly worse; the four ties are DeiT-S 65\% and DeiT-B 70\%, 65\% and 50\%, where the differences are
$-0.17$ to $+0.30$ points. Across the eight settings at budgets of 55\%--40\%, it recovers 98--149\% of the IQP's
advantage over the ILP, computed as $100\,(a_\text{hybrid}-a_\text{ILP})/(a_\text{IQP}-a_\text{ILP})$. It is also
significantly better than equal-cost random sampling in 11 settings and evolution in 9, with one significant loss
(DeiT-B, 40\%, $-1.63$ points). On ImageNet-100 the hybrid and evolution were mostly not significantly different on val100 but the hybrid was clearly ahead on
val1k (Table~\ref{tab:baselines}). With 1000-class fine-tuning that advantage now appears on the main metric.
On DeiT-S, random and evolution often keep the lin start, as on ImageNet-100, which explains the ties at 55\% and 50\%.

\subsection{A second architecture: Swin-T}
\label{sec:swin}
DeiT-S and DeiT-B share one design: twelve identical blocks with global attention. To test whether the findings depend
on it, we repeated the full ImageNet-1k pipeline of Section~\ref{sec:in1k} on Swin-T~\cite{liu2021swin}
(\texttt{swin\_tiny\_patch4\_window7\_224} from timm, 81.36\% top-1 on the 50{,}000 validation images). Swin-T is
hierarchical: its twelve blocks form four stages of 2, 2, 6 and 2 blocks at resolutions $56^2$, $28^2$, $14^2$ and
$7^2$, and it uses shifted-window attention with a relative position bias. The block library, scores, budgets,
searches and fine-tuning recipe are unchanged. Two adaptations are needed. Head pruning also prunes the corresponding
columns of the relative-position-bias table. Because block latency now differs by layer (from 0.50\,ms in the last
stage to 3.12\,ms in the first), the latency table and all searches use per-layer latencies; the summed table predicts
the end-to-end latency of 20 random structures with a mean absolute relative error of 0.7\% (maximum 2.0\%).

\paragraph{Before fine-tuning.}
Additivity is somewhat weaker than on DeiT. On the 400 random structures the plain sum ranks with Kendall $\tau=0.855$,
raised to 0.937 by the pair terms; 97\% of the structures are super-additive (median act./add 1.22), and the mean
relative pair term is 0.27 at layer distance $\le2$ and 0.04 beyond. The metric share of the mean KL additivity error
is 43\% (structures) and 32\% (pairs), between the two DeiT models. The additive search again breaks down: before
fine-tuning, the IQP is ahead by 5.3, 16.6 and 6.9 points at 50\%, 45\% and 40\% (Table~\ref{tab:swin}). The
transition lies between 55\% and 50\%, one budget step later than on DeiT.

\begin{table*}[t]
\centering
\caption{Swin-T on full ImageNet-1k: top-1 (\%) on the 50{,}000 validation images before and after the 2{,}000-step
fine-tune (after: mean of three seeds). Last column: IQP $-$ ILP after fine-tuning with 95\% paired bootstrap interval
(\sig: interval excludes 0). The parent reaches 81.36\%.}
\label{tab:swin}
\small
\setlength{\tabcolsep}{4pt}
\begin{tabular}{lcccccccc}
\toprule
 & \multicolumn{2}{c}{before fine-tuning} & \multicolumn{5}{c}{after fine-tuning} & \\
\cmidrule(lr){2-3}\cmidrule(lr){4-8}
budget & ILP & IQP & ILP & IQP & lin & random / evol. & hybrid & IQP $-$ ILP [95\% CI]\\
\midrule
70\% & 78.64 & 79.13 & 80.36 & 80.25 & 80.26 & 80.26 / 80.26 & 80.32 & $-0.12$ [$-0.23$, $0.00$]\\
65\% & 78.08 & 77.99 & 80.32 & 79.79 & 80.30 & 80.30 / 80.30 & 80.44 & $-0.54$ [$-0.68$, $-0.40$]\sig\\
60\% & 76.88 & 77.54 & 79.98 & 80.04 & 79.98 & 79.98 / 79.98 & 80.09 & $+0.05$ [$-0.06$, $+0.16$]\\
55\% & 75.32 & 75.88 & 79.45 & 79.48 & 79.45 & 79.45 / 79.35 & 79.14 & $+0.03$ [$-0.08$, $+0.15$]\\
50\% & 68.99 & 74.25 & 77.92 & 78.45 & 78.25 & 78.25 / 78.25 & 78.46 & $+0.52$ [$+0.36$, $+0.69$]\sig\\
45\% & 53.57 & 70.18 & 75.79 & 77.01 & 77.01 & 77.01 / 77.01 & 76.94 & $+1.21$ [$+1.02$, $+1.40$]\sig\\
40\% & 54.90 & 61.76 & 74.04 & 74.10 & 74.78 & 74.78 / 74.78 & 75.12 & $+0.06$ [$-0.14$, $+0.26$]\\
\bottomrule
\end{tabular}
\end{table*}

\paragraph{After fine-tuning.}
Unlike on DeiT, fine-tuning reduces the gap to a small value. After 2{,}000 steps the IQP is ahead by at most 1.21 points, and the
difference is significant in only three of seven settings, one of them in favor of the ILP (65\%, $-0.54$). For
comparison, the gap at 45\% shrinks from 16.6 to 1.2 points on Swin-T, but only from 31.9 to 6.5 points on DeiT-S and
from 37.4 to 16.1 points on DeiT-B. The search method matters correspondingly little. All six searches end within
1.3 points of each other at every budget; lin, random and evolution select the same structure at most budgets. The
hybrid is significantly better than the ILP in 3, not significantly different in 3 and worse in 1 of the 7 settings
(2/4/1 against the IQP); its only loss is at 55\% ($-0.31$ against the ILP).

\subsection{A language model: Qwen2.5-0.5B}
\label{sec:llm}
Puzzle~\cite{bercovich2024puzzle} applies blockwise replacement search to large language models. To test whether our
findings carry over to that setting, we repeated the study on Qwen2.5-0.5B~\cite{yang2024qwen25}, a decoder-only
transformer with 24 layers, grouped-query attention with 14 query and 2 key--value heads, and a SwiGLU MLP.

\paragraph{Setup.}
The library has the same seven options per layer. Head pruning keeps one of the two key--value groups and its seven
query heads; the MLP options prune or factorize the SwiGLU intermediate channels. Candidates are distilled with the
loss of Section~\ref{sec:setup} for 3{,}000 steps (batch of eight 512-token sequences, learning rate
$3\times10^{-4}$) on FineWeb-Edu~\cite{penedo2024fineweb}. \emph{calib} consists of 128 held-out 512-token sequences
from separate documents (65{,}408 next-token predictions), and every score is the mean KL over the full
151{,}936-token vocabulary. We measure all $24\times7=168$ single replacements, 4{,}053 pairs chosen with the design of
Section~\ref{sec:setup} (all option pairs at layer distance $\le2$, eight per farther layer pair), and 400 random
structures with 2--6 replaced layers. The latency table is measured for a prefill of $8\times512$ tokens in bf16 on the
RTX 5090. Removing a layer saves almost all of its latency, removing attention 43\% and removing the MLP 51\%, whereas
halving the heads saves only 3\% because the key--value projections are already small. The summed table predicts the
end-to-end latency of 20 random structures with a mean absolute relative error of 4.1\% (maximum 8.3\%). Because the
model is small and already compact, we use budgets from 90\% to 60\%. The four searches (ILP, IQP, lin and hybrid) are
unchanged except that lin and the hybrid's ranking use a bucketed distribution, because storing full-vocabulary logit
changes for all 168 single replacements is impractical. On the first 16 \emph{calib} sequences, the parent's 256 most
probable tokens are fixed per position; for the parent and for every single replacement we store 257 bucket logits per
position, the logits of these 256 tokens and the log-sum-exp of all other logits. lin then sums the single-replacement
changes of these 257 bucket logits and evaluates the KL between the resulting 257-way distributions. Summing changes
of the rest bucket is an additional approximation, since the log-sum-exp is not linear; the decomposition below is
computed on the same bucketed distribution, whereas the searches and all other scores use the full vocabulary. The hybrid measures on every fourth
\emph{calib} sequence. Each child is fine-tuned by global distillation from the parent (KL over the full vocabulary,
2{,}000 steps of $16\times512$ tokens, AdamW, learning rate $5\times10^{-5}$, weight decay 0.05, 100 warm-up steps,
cosine decay; two seeds). We report the macro-average of zero-shot accuracy on ARC-Easy and ARC-Challenge (test
sets)~\cite{clark2018arc}, HellaSwag~\cite{zellers2019hellaswag}, PIQA~\cite{bisk2020piqa} and
WinoGrande~\cite{sakaguchi2020winogrande} (validation sets; length-normalized log-likelihood, partial scoring for
WinoGrande), and the perplexity on WikiText-2~\cite{merity2017wikitext} (test set, 512-token chunks). The parent
reaches 54.15\% and a perplexity of 16.90. Confidence intervals resample examples within each task (paired, seeds
averaged per example) and are macro-averaged.

\paragraph{Before fine-tuning.}
On random structures with 2--6 of the 24 layers replaced, the scores are nearly additive: the plain sum ranks with
Kendall $\tau=0.964$ (0.975 with pair terms), only 48\% of the structures are super-additive, the median act./add ratio
is 1.00, and the mean relative pair term is 0.057 at layer distance $\le2$ and 0.023 beyond. Section~\ref{sec:q2} analyzes the decomposition of the error on this model.

The searched structures, however, replace 17--23 of the 24 layers at
70\%--60\%, far beyond the random structures. There additivity fails: the calibration KL of the ILP's choice exceeds
the IQP's by 33\%, 44\% and 47\% at 70\%, 65\% and 60\%. The ILP meets these budgets by removing the attention of 5, 7
and 9 adjacent layers; the IQP's longest run of one option is 3 layers.

\begin{table*}[t]
\centering
\caption{Qwen2.5-0.5B after fine-tuning (two seeds): macro-averaged zero-shot accuracy over five tasks (\%) and
WikiText-2 perplexity. IQP $-$ ILP with 95\% paired bootstrap interval (\sig: excludes 0). The parent reaches 54.15\%
and perplexity 16.90.}
\label{tab:llm}
\small
\setlength{\tabcolsep}{3.5pt}
\begin{tabular}{lccccccccc}
\toprule
 & \multicolumn{5}{c}{zero-shot accuracy} & \multicolumn{4}{c}{perplexity}\\
\cmidrule(lr){2-6}\cmidrule(lr){7-10}
budget & ILP & IQP & lin & hybrid & IQP $-$ ILP [95\% CI] & ILP & IQP & lin & hybrid\\
\midrule
90\% & 52.96 & 52.81 & 53.49 & 53.83 & $-0.15$ [$-0.69$, $+0.38$] & 18.8 & 18.7 & 19.2 & 19.1\\
85\% & 52.17 & 52.95 & 53.21 & 53.23 & $+0.78$ [$+0.24$, $+1.33$]\sig & 19.8 & 19.5 & 20.6 & 20.0\\
80\% & 51.65 & 51.67 & 52.49 & 51.72 & $+0.03$ [$-0.60$, $+0.64$] & 21.7 & 20.9 & 22.1 & 21.6\\
75\% & 50.07 & 50.33 & 50.78 & 50.93 & $+0.26$ [$-0.34$, $+0.85$] & 23.4 & 23.4 & 24.2 & 23.0\\
70\% & 49.01 & 48.67 & 48.17 & 49.74 & $-0.35$ [$-1.01$, $+0.32$] & 27.6 & 24.2 & 29.4 & 25.1\\
65\% & 46.72 & 48.14 & 46.75 & 47.79 & $+1.43$ [$+0.75$, $+2.08$]\sig & 34.3 & 28.2 & 51.0 & 28.8\\
60\% & 45.39 & 46.63 & 44.66 & 46.46 & $+1.24$ [$+0.53$, $+1.95$]\sig & 60.9 & 32.5 & 47.2 & 57.6\\
\bottomrule
\end{tabular}
\end{table*}

\paragraph{After fine-tuning.}
Table~\ref{tab:llm} shows the same qualitative transition as on the vision models, on a different scale. Down to
70\% the four searches are within about one point of zero-shot accuracy, and IQP $-$ ILP is significant only at 85\%
($+0.78$). At 65\% and 60\% the IQP is significantly ahead, by 1.43 and 1.24 points. Perplexity separates the searches
more clearly and earlier: the ILP children reach 27.6, 34.3 and 60.9 at 70\%, 65\% and 60\%, against 24.2, 28.2 and
32.5 for the IQP (seed ranges below 0.6). The zero-shot accuracies of a 0.5B model are close to chance on some tasks
(ARC-Challenge 31.9\% for the parent), which compresses all differences; we therefore regard perplexity as the more
sensitive measure here. Per task, after Holm adjustment over the 70 task-level comparisons with the ILP, 12 are
significant, all in favor of the IQP or the hybrid and all on ARC-Easy and HellaSwag, half of them at 65\% and 60\%;
the near-chance tasks show no significant difference. As on DeiT, the additive children start much further behind (perplexity 217 against 97 at 60\%)
and do not catch up within the fine-tuning budget.

The cheap repairs behave as the decomposition suggests. Because the summed logits overshoot, lin is not a reliable
ranker on this model: its children have the highest perplexity at 70\% and 65\%. The hybrid, which only uses lin to
propose moves and judges them by measured KL, is significantly better than the ILP in zero-shot accuracy in 6 of 7
settings (no significant difference at 80\%) and not significantly different from the IQP in 5 (better in 2). Its
perplexity is close to the IQP's except at 60\%. There it starts from the ILP's solution, finds a single improving
move among the moves proposed by lin, and stops in a configuration that still removes the attention of nine adjacent
layers (perplexity 57.6). A hybrid can only be as good as the neighborhood its ranker proposes; on this model the
ranker does not propose the moves that break up the long run.

\subsection{Run-length constraints}
\label{sec:runlen}
We applied the run-length constraints of Section~\ref{sec:runlen_method} on full ImageNet-1k, re-running the ILP and
IQP searches with the existing scores and latency tables and fine-tuning the new structures with the recipe of
Section~\ref{sec:setup} (three seeds). All constrained programs were solved to optimality.

\begin{table*}[t]
\centering
\caption{Run-length interventions on full ImageNet-1k (2{,}000-step fine-tune, three seeds). Entries are IQP $-$ ILP
top-1 differences (points). ``Run'': longest run of adjacent layers with a removal (\texttt{attn\_skip},
\texttt{mlp\_skip}, \texttt{block\_skip}) in the unconstrained ILP / IQP structures. Last two columns: top-1 of ILP and
IQP under the removal-class cap $R=2$. Every difference is significant after Holm
adjustment over the 39 intervention comparisons, and its three per-seed differences agree in sign (per-seed spread at
most 0.4 points).}
\label{tab:runlen}
\small
\setlength{\tabcolsep}{4pt}
\begin{tabular}{llccccccccc}
\toprule
 & & \multicolumn{2}{c}{unconstrained} & \multicolumn{2}{c}{same option} & \multicolumn{2}{c}{removal class} & \multicolumn{2}{c}{removal class, $R=2$}\\
\cmidrule(lr){3-4}\cmidrule(lr){5-6}\cmidrule(lr){7-8}\cmidrule(lr){9-10}
model & budget & gap & run & $R=3$ & $R=2$ & $R=3$ & $R=2$ & ILP & IQP\\
\midrule
\multirow{4}{*}{DeiT-S}
 & 55\% & $+7.03$ & 4 / 3 & $+7.03$ & $+7.03$ & $+5.89$ & $\mathbf{+0.74}$ & 70.21 & 70.95\\
 & 50\% & $+9.66$ & 5 / 3 & $+9.66$ & $+7.61$ & $+6.75$ & $\mathbf{+1.13}$ & 65.48 & 66.61\\
 & 45\% & $+6.54$ & 3 / 4 & $+6.54$ & $+8.47$ & $+6.46$ & $\mathbf{+2.07}$ & 59.87 & 61.94\\
 & 40\% & $+17.50$ & 4 / 5 & $+17.50$ & $+13.31$ & $+12.82$ & $\mathbf{+4.03}$ & 54.91 & 58.94\\
\midrule
\multirow{4}{*}{DeiT-B}
 & 55\% & $+5.79$ & 5 / 4 & $+4.64$ & $+4.38$ & $+2.79$ & $\mathbf{+1.82}$ & 73.69 & 75.51\\
 & 50\% & $+8.32$ & 5 / 4 & $+11.05$ & $+7.95$ & $+3.58$ & $\mathbf{+5.21}$ & 69.28 & 74.49\\
 & 45\% & $+16.10$ & 6 / 6 & $+10.34$ & $+6.36$ & $+5.57$ & $\mathbf{+4.51}$ & 64.65 & 69.16\\
 & 40\% & $+14.62$ & 6 / 5 & $+15.11$ & $+5.32$ & $+10.08$ & $\mathbf{+2.63}$ & 63.01 & 65.64\\
\bottomrule
\end{tabular}
\end{table*}

Table~\ref{tab:runlen} gives a direct answer. Capping runs of one option is weak: at $R=3$ the DeiT-S structures are
unchanged, and at $R=2$ the gap shrinks by more than five points only at DeiT-B 45\% and 40\%; elsewhere the ILP already satisfies the cap or alternates similar removals. The removal-class
cap with $R=2$ shrinks the gap at all eight settings, from 5.8--17.5 points to 0.7--5.2 points; on DeiT-S at 55\% and
50\% it nearly vanishes. The reduction comes almost entirely from the additive search: the ILP gains 3.2--13.8 points
over its unconstrained structures (for example from 42.96\% to 54.91\% on DeiT-S at 40\%, and from 49.21\% to 63.01\%
on DeiT-B at 40\%), whereas the IQP changes by between $-1.9$ and $+1.8$ points. Allowing three adjacent removals
($R=3$) brings back most of the gap at 40\%. Long runs of adjacent removals, which an additive objective does not
penalize, are therefore a major failure pattern behind the post-fine-tuning gap on DeiT in the regime studied. The
re-solved programs may also change the number, type and position of the removals, so this is not a matched
intervention and does not isolate adjacency alone. Nor is it the only factor: 1.8--5.2 points remain on DeiT-B, and the remaining gap is not monotone in $R$ (DeiT-B 50\%: 3.6 points at
$R=3$, 5.2 at $R=2$). A practical consequence is that a single family of linear constraints brings the additive ILP
within 0.7--5.2 points of the IQP, without any pair measurements.

\paragraph{Forbidding whole-block removal.}
An earlier, coarser intervention removed \texttt{block\_skip} from the library. At 40\% the restricted problem is
infeasible for DeiT-S.

Table~\ref{tab:noskip} gives the results; they support the explanation only in part. Forbidding whole-block removal shrinks the DeiT-B gap at
45\% and 40\% from 15--16 points to 0--5 points, but six of the seven restricted settings remain significant. Deprived
of \texttt{block\_skip}, the additive ILP stacks the next-cheapest option: it removes the MLP of layers 1--7 on DeiT-B
at 55\%--45\% and of layers 2--10 on DeiT-S at 45\%, whereas the IQP interleaves options. At 40\% on DeiT-B the budget
forces both searches to remove the MLP of eleven consecutive layers, and the gap vanishes.

\subsection{Does more fine-tuning close the gap?}
\label{sec:curve}
The 2{,}000-step fine-tune is about a fifth of an ImageNet-1k epoch, so the large DeiT gaps might only reflect a
short schedule. We therefore fine-tuned the ILP and IQP children of the three most aggressive budgets for 10{,}000
steps (two seeds; recipe of Section~\ref{sec:setup} with cosine decay over 10{,}000 steps), on DeiT also the hybrid's
children, and evaluated them at 0,
500, 1{,}000, 2{,}000, 5{,}000 and 10{,}000 steps; values at step 2{,}000 differ from Table~\ref{tab:in1k} because the
learning rate has not yet decayed there.

\begin{table*}[t]
\centering
\caption{Fine-tuning recovery on full ImageNet-1k: IQP $-$ ILP top-1 difference (percentage points; two seeds; 95\%
paired bootstrap interval) during a 10{,}000-step fine-tune. \sig: interval excludes 0. After 10{,}000 steps:
seed-averaged top-1 of ILP / IQP (/ hybrid on DeiT) and the hybrid's difference to the IQP (\sig: significant after
Holm over the six DeiT tests).}
\label{tab:curve}
\small
\setlength{\tabcolsep}{4pt}
\begin{tabular}{llccc}
\toprule
budget & step & DeiT-S & DeiT-B & Swin-T\\
\midrule
\multirow{6}{*}{50\%} & 0 & $+38.11$\sig & $+45.15$\sig & $+5.25$\sig\\
 & 500 & $+13.46$\sig & $+11.80$\sig & $+0.61$\sig\\
 & 2{,}000 & $+8.45$\sig & $+7.23$\sig & $+0.56$\sig\\
 & 10{,}000 & $+4.64$ [$+4.35$, $+4.94$]\sig & $+4.67$ [$+4.41$, $+4.93$]\sig & $+0.30$ [$+0.14$, $+0.47$]\sig\\
 & top-1 & 66.64 / 71.28 / 72.34 & 71.88 / 76.55 / 76.34 & 78.77 / 79.06\\
 & hyb$-$IQP & $+1.06$\sig & $-0.21$ & --\\
\midrule
\multirow{6}{*}{45\%} & 0 & $+31.87$\sig & $+37.46$\sig & $+16.61$\sig\\
 & 500 & $+8.41$\sig & $+22.58$\sig & $+1.75$\sig\\
 & 2{,}000 & $+5.33$\sig & $+13.40$\sig & $+1.27$\sig\\
 & 10{,}000 & $+3.34$ [$+3.02$, $+3.64$]\sig & $+9.06$ [$+8.74$, $+9.38$]\sig & $+1.26$ [$+1.07$, $+1.46$]\sig\\
 & top-1 & 64.61 / 67.95 / 69.98 & 65.34 / 74.40 / 75.27 & 76.84 / 78.10\\
 & hyb$-$IQP & $+2.03$\sig & $+0.87$\sig & --\\
\midrule
\multirow{6}{*}{40\%} & 0 & $+40.54$\sig & $+9.53$\sig & $+6.86$\sig\\
 & 500 & $+22.40$\sig & $+19.23$\sig & $+0.24$\sig\\
 & 2{,}000 & $+14.85$\sig & $+12.23$\sig & $+0.18$\\
 & 10{,}000 & $+9.87$ [$+9.52$, $+10.22$]\sig & $+8.01$ [$+7.69$, $+8.33$]\sig & $0.00$ [$-0.20$, $+0.19$]\\
 & top-1 & 55.88 / 65.75 / 67.08 & 61.57 / 69.57 / 70.93 & 75.87 / 75.87\\
 & hyb$-$IQP & $+1.33$\sig & $+1.36$\sig & --\\
\bottomrule
\end{tabular}
\end{table*}

\begin{figure*}[t]
\centering
\includegraphics{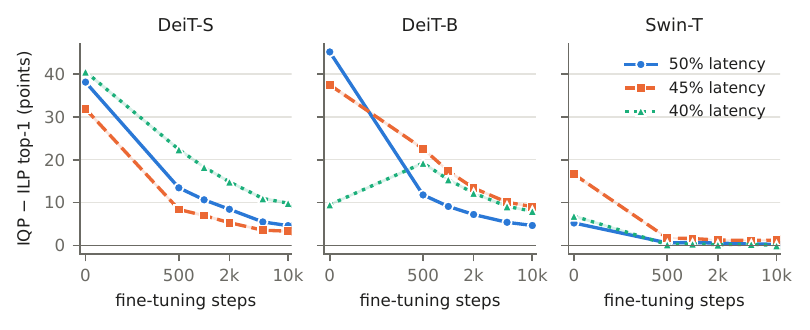}
\caption{Fine-tuning recovery on full ImageNet-1k: IQP $-$ ILP top-1 difference during a 10{,}000-step fine-tune
(mean of two seeds; shaded: 95\% paired bootstrap interval, at most 1 point wide and therefore barely visible). The
step axis is linear up to 500 and logarithmic beyond. On DeiT the gap shrinks slowly and stays at 3--10 points; on
Swin-T it shrinks to at most about 1.8 points within 500 steps.}
\label{fig:curve}
\end{figure*}

Table~\ref{tab:curve} and Figure~\ref{fig:curve} answer the objection. On Swin-T the gap shrinks within the first 500 steps, from 5.3--16.6
points to at most 1.8 points, and then stays essentially flat (0.0--1.3 points at 10{,}000 steps). On DeiT-S and
DeiT-B the gap also shrinks, but slowly: after five times the fine-tuning of the main recipe it is still 3.3--9.9
points, and between 5{,}000 and 10{,}000 steps it shrinks by only 0.2--1.1 points. Even after 10{,}000 steps the DeiT gaps are
larger than the Swin-T gaps after 500 steps. The difference between the architectures is therefore not explained by the
fine-tuning length alone; within schedules of up to 10{,}000 steps, the damage done by the additive choice on DeiT is
reduced but not removed by a longer schedule. The hybrid's lead is not an artefact of the short schedule either: after 10{,}000 steps
it is 4.5--11.2 points ahead of the ILP and 0.9--2.0 points ahead of the IQP in five of the six DeiT settings
(significant after Holm), and on par with the IQP at DeiT-B 50\% ($-0.21$), as after 2{,}000 steps.

\begin{table}[t]
\centering
\caption{Learning-rate sensitivity on the evaluation set (full ImageNet-1k, 2{,}000 steps, three seeds). Each method
at the rate with the best evaluation-set accuracy in $\{0.2,0.5,1,2\}\times10^{-4}$ (in parentheses, units of
$10^{-4}$; main recipe 0.5); the ILP was also tried at $4\times10^{-4}$ because its best rate lay at the edge of the
grid. Since the rates are selected on the evaluation set, these are not held-out estimates. Last columns: IQP $-$ ILP and
hybrid $-$ IQP at the best rates; main-recipe IQP $-$ ILP for comparison. S, B: DeiT-S, DeiT-B. \sig: significant after Holm over the 12
comparisons, paired over images.}
\label{tab:lr}
\footnotesize
\setlength{\tabcolsep}{2pt}
\begin{tabular}{@{}llccccrr@{}}
\toprule
 & budget & ILP & IQP & hybrid & IQP$-$ILP & hyb$-$IQP & main\\
\midrule
S & 50\% & 62.40\,{\scriptsize(2)} & 68.60\,{\scriptsize(2)} & 69.85\,{\scriptsize(1)} & $+6.20$\sig & $+1.25$\sig & $+9.66$\\
 & 40\% & 51.72\,{\scriptsize(4)} & 62.41\,{\scriptsize(2)} & 63.92\,{\scriptsize(2)} & $+10.69$\sig & $+1.51$\sig & $+17.50$\\
B & 50\% & 68.72\,{\scriptsize(2)} & 74.68\,{\scriptsize(1)} & 74.39\,{\scriptsize(1)} & $+5.97$\sig & $-0.30$\sig & $+8.32$\\
 & 40\% & 56.43\,{\scriptsize(2)} & 65.91\,{\scriptsize(2)} & 67.49\,{\scriptsize(2)} & $+9.48$\sig & $+1.58$\sig & $+14.62$\\
\bottomrule
\end{tabular}
\end{table}

\paragraph{Learning rate.} All searches above share one fine-tuning recipe, which might suit some children better than
others. We therefore swept the learning rate for the ILP, IQP and hybrid children at 50\% and 40\% on DeiT-S and DeiT-B
(Table~\ref{tab:lr}). There is no separate validation split, so each rate is selected on the evaluation set and the
results measure sensitivity to the learning rate rather than held-out accuracy. The additive children profit most from a larger rate (DeiT-B 40\%: 49.2 at $5\times10^{-5}$,
56.4 at $2\times10^{-4}$), so tuning narrows the gap by 2.3--6.8 points. It does not close it: with every method at its
own best rate the IQP is still 6.0--10.7 points and the hybrid 5.7--12.2 points ahead of the ILP. The ILP's optimum
lies inside the grid on DeiT-B and at 50\% on DeiT-S ($4\times10^{-4}$ is worse); at DeiT-S 40\% it is at the edge,
but the gain from $2$ to $4\times10^{-4}$ is only 1.1 points. The hybrid stays ahead of the IQP at 40\% (by 1.5--1.6
points) and DeiT-S 50\%, and is 0.3 points behind at DeiT-B 50\%, as with the main recipe.

\subsection{Search-seed variance and larger evaluation budgets}
\label{sec:seeds}
The equal-cost baselines of Sections~\ref{sec:q1} and~\ref{sec:in1k} were run with one seed and exactly the
hybrid's number of real evaluations. We therefore repeated the random and evolutionary searches on full ImageNet-1k
with three seeds, both with the hybrid's number of evaluations ($1\times$, 74--243 structures per budget) and with ten
times as many ($10\times$). Starting points, subset, and evolution settings are those of Section~\ref{sec:setup};
seed 0 at $1\times$ reproduces the earlier runs exactly. Table~\ref{tab:seeds} reports the top-1 accuracy of each
search's best structure before fine-tuning.

\paragraph{Before fine-tuning.}
On DeiT-S the hybrid's structure is the most accurate at every budget, also against evolution with ten times the
evaluations (by 0.4--2.9 points on average). On DeiT-B the hybrid is on par with or ahead of evolution at $10\times$
down to 60\%, but at 50\%--40\% black-box search finds far more accurate structures before fine-tuning: equal-cost
random sampling is ahead by 9--17 points at 45\%--40\%, and evolution at $10\times$ by 5--33 points. More evaluations
do not always help on the full validation set, since the searches optimize KL on a 1{,}000-image subset (DeiT-S, 40\%:
random $10\times$ is worse than random $1\times$), and the earlier fine-tuned comparison (Table~\ref{tab:in1k}) already
showed the hybrid ahead of equal-cost evolution at DeiT-B 50\% and 45\% after fine-tuning, although evolution was far
ahead before.

\paragraph{After fine-tuning.}
To settle the DeiT-B case, we fine-tuned the best structure of each $10\times$ evolution seed at 50\%--40\% with the
recipe of Section~\ref{sec:setup} (Table~\ref{tab:evo10}). At 50\% and 45\% the large pre-fine-tuning lead of
evolution disappears: with ten times the hybrid's real evaluations, evolution ends 0.17--0.27 points ahead, less than
the spread of about one point between its own search seeds. At 40\% evolution keeps a clear lead of 2.3--4.1 points in
every seed. This exposes limits of the current approximation and search strategy: at 40\% latency on DeiT-B, 10--12 of the
twelve blocks are replaced in the structures compared, and a search that measures many complete structures finds
better ones than the pairwise model and the hybrid's local moves. Interactions of more than two blocks are one
possible reason; we did not measure them. These fine-tuned comparisons cover only three DeiT-B budgets; for the
other settings Table~\ref{tab:seeds} compares the searches before fine-tuning only.

\section{Analysis}
\label{sec:analysis}
\subsection{Where does the additivity error come from?}
\label{sec:q2}
\begin{table}[t]
\centering
\caption{Decomposition of the additivity error (\emph{calib}): measured pairs and 400 random structures (struct.). ``Metric'': mean metric part. Components
are signed and averaged over the pairs or structures; the metric share is the mean metric part divided by the mean of
act $-$ add, so it is not a decomposition of an absolute error and need not lie in [0, 1] in general. ``Acc.''
uses the 100-way top-1 drop (percentage points) instead of KL as the score.}
\label{tab:decomp}
\footnotesize
\setlength{\tabcolsep}{2.5pt}
\begin{tabular}{llccccc}
\toprule
model & set & $\tau$ add & $\tau$ lin & act$-$add & metric & share\\
\midrule
\multirow{4}{*}{DeiT-S} & pairs, KL & 0.924 & 0.931 & 0.0171 & 0.0102 & 59\%\\
 & struct., KL & 0.894 & 0.911 & 0.0694 & 0.0454 & 65\%\\
 & pairs, acc. & 0.811 & 0.875 & 0.255 & 0.094 & 37\%\\
 & struct., acc. & 0.825 & 0.897 & 1.256 & 0.446 & 36\%\\
\midrule
\multirow{4}{*}{DeiT-B} & pairs, KL & 0.934 & 0.952 & 0.0143 & 0.0038 & 27\%\\
 & struct., KL & 0.894 & 0.932 & 0.1095 & 0.0267 & 24\%\\
 & pairs, acc. & 0.820 & 0.882 & 0.195 & 0.034 & 17\%\\
 & struct., acc. & 0.781 & 0.874 & 1.748 & 0.282 & 16\%\\
\bottomrule
\end{tabular}
\end{table}

\paragraph{Vision models.}
Table~\ref{tab:decomp} shows a clear difference between the two models.
On DeiT-S, 59--65\% of the mean KL additivity error is metric: much of what an interaction-aware search ``learns''
from pair measurements is curvature of the score rather than functional interaction. On DeiT-B, the metric share is
only 24--27\%, and most of the mean error of the wider model is attributed to the functional component once several
blocks are replaced.
In both models the lin predictor ranks structures better than the plain sum (higher $\tau$). This holds even for the
accuracy-drop score, where the metric share is smaller. On full ImageNet-1k the metric share is 63--67\% on DeiT-S and 34--36\% on DeiT-B (Section~\ref{sec:in1k}), and
32--43\% on Swin-T (Section~\ref{sec:swin}). The difference between the DeiT models matches the behavior of the
cheap repairs: lin is competitive with the IQP on DeiT-S but falls behind on DeiT-B at aggressive budgets
(Sections~\ref{sec:q1} and~\ref{sec:in1k}).

\paragraph{Validity of the second-order reading.}
Section~\ref{sec:decomp} bounds the remainder of the quadratic reading by $\phi(\pm\omega)$. In practice $\omega$ is large even for small
structures, because a few logits move by several nats, so the bound is not informative and we measured the remainder
directly on full ImageNet-1k (\emph{calib}; 80 random structures with 2--6 replaced blocks and, at each budget from 70\% to 40\%, 104 structures: the 60 band
and 40 near structures of Section~\ref{sec:rankq} and the final structures of the ILP, IQP, lin and hybrid, of which
two coincide at DeiT-B 70\%). For the small structures the exact summed-logit KL is 1.07
(DeiT-S) and 1.11 (DeiT-B) times its quadratic term (medians), and the quadratic cross term recovers 81\% and 73\% of
the exact metric part. At the budgets searched, the KL exceeds its quadratic term by a factor of 1.2--1.4 (DeiT-S) and
1.2--1.8 (DeiT-B), and the cross term recovers only 64--73\% and 44--74\% of the metric part, although it stays highly
correlated with it (Pearson 0.87--0.94). The cross-term reading is therefore quantitatively accurate only for small
perturbations and qualitative at aggressive budgets. None of the reported numbers depend on the expansion: lin evaluates the summed-logit
KL exactly, from the cached logits of the 84 single-block runs, and subtracting add gives the exact metric part.

\paragraph{Language model.}
The decomposition is
unlike that of the vision models. On the bucketed distribution, the summed-logit prediction overshoots: the metric part
exceeds the whole additivity error (139\% on pairs, 225\% on structures), that is, the functional part is negative:
the KL of the actual child is below that of the summed-logit prediction. This is compatible with partial cancellation
of the logit changes, but a negative functional part does not establish it, since different logits can give the same
KL. It is not an artifact of the bucketing: on four \emph{calib} sequences (2{,}044 positions) we recomputed act, add
and lin on the full 151{,}936-token vocabulary for 60 random structures and the 28 searched structures. The summed
logits overshoot for 95--100\% of the structures on the full vocabulary as well (92--96\% bucketed), the metric share
is even larger (226--326\% against 203--239\%), and the bucketed lin orders the structures almost exactly as the
full-vocabulary lin (Kendall $\tau$ 0.993 on the random and 1.000 on the searched structures). Accordingly, lin ranks the random structures \emph{worse} than the plain sum
($\tau=0.902$ against 0.959 on the bucketed KL).

\subsection{Why does fine-tuning repair the damage on Swin-T but not on DeiT?}
\label{sec:why}
\begin{table}[t]
\centering
\caption{How the searches meet the aggressive budgets (55\%--40\%, full ImageNet-1k): share of the saved block latency
that comes from removing whole blocks or only the attention, and the longest run of adjacent removed blocks (range
over the four budgets).}
\label{tab:swin_why}
\footnotesize
\setlength{\tabcolsep}{2.5pt}
\begin{tabular}{llccc}
\toprule
model & search & whole block & attention & longest run\\
\midrule
DeiT-S & ILP & 49--75\% & 6--33\% & 2--3\\
       & IQP & 55--66\% & 7--13\% & 1--2\\
DeiT-B & ILP & 82--92\% & 0\% & 5--6\\
       & IQP & 50--69\% & 8--14\% & 1--3\\
Swin-T & ILP & 0--18\% & 78--98\% & 0--1\\
       & IQP & 0--35\% & 58--93\% & 0--1\\
\bottomrule
\end{tabular}
\end{table}

Table~\ref{tab:swin_why} suggests an explanation in terms of how the budget is met. In DeiT every block has the same
latency, and removing a whole block is the cheapest way to save it. Residual networks tolerate the removal of single
blocks while the error grows as more of them are removed~\cite{veit2016residual}. The additive ILP therefore removes many whole
blocks, on DeiT-B five or six adjacent ones, exactly the configurations whose pair terms the IQP penalizes; its
children start from 3--16\% top-1 at 45\% and do not recover within 2{,}000 steps. In Swin-T, latency is concentrated
in the attention of the high-resolution early stages (attention is 61\% of a block's latency on average). Both searches
meet the budget mainly by removing attention there while keeping the MLP and the residual path, and they rarely remove
whole blocks. The resulting additivity errors are large before fine-tuning but are largely recovered by fine-tuning: the
additive child at 45\% starts from 53.6\% and ends within 1.2 points of the IQP's. The run-length experiments (Section~\ref{sec:runlen})
support this explanation: capping runs of adjacent removals removes most of the DeiT gap, almost entirely by
improving the additive solution. The fine-tuning curves (Section~\ref{sec:curve}) show that the contrast between the
architectures is not an artifact of the fine-tuning length.

Removing an option changes the search space and the replacement types at the same time, so this intervention alone
cannot isolate adjacency; it is consistent with the run-length results above, in which the library is unchanged and
the gap shrinks once removals may no longer occupy more than two adjacent layers. Together, the two interventions
suggest that adjacent removals, not whole-block removal as such, are what the additive search gets wrong on DeiT.
On Swin-T the runs consist of removed windowed attention with the MLP and residual path intact, and the gap is largely
recovered by fine-tuning (Table~\ref{tab:curve}). Which replacements fine-tuning can recover is a property of the
architecture that this study can describe but not predict. Su et al.\ report that with a linear repair the optimal
set of removed language-model layers becomes a single contiguous block~\cite{su2026repair}, whereas our run-length
constraints forbid long runs. The two findings are compatible: in both settings adjacent removals interact strongly
without repair, and whether a repair step or fine-tuning absorbs this interaction depends on the setting, as the
contrast between DeiT and Swin-T shows.

\subsection{How well do the predictors rank where the search decides?}
\label{sec:rankq}
The rank correlations of Table~\ref{tab:additivity} are computed on random structures with 2--6 replaced blocks, far
from the structures that the searches compare at aggressive budgets. On full ImageNet-1k we therefore sampled, at each
budget from 70\% to 40\%, 60 uniformly random feasible structures whose latency lies within five points below the
budget (\emph{band}) and 40 one- and two-layer neighbors of the hybrid's final structure in the same latency range
(\emph{near}), the region in which the local searches make their decisions, and measured their calibration KL.
Table~\ref{tab:rankq} reports Kendall $\tau$ between each predictor and the measured KL.

\begin{table}[t]
\centering
\caption{Kendall $\tau$ between predicted and measured calibration KL inside each budget (full ImageNet-1k): add (sum
of single scores, ILP objective), pair (with measured pair terms, IQP objective) and lin. Band: 60 random feasible
structures within five points below the budget; near: 40 neighbors of the hybrid's final structure. Random structures
with 2--6 replaced blocks for reference: 0.88 / 0.95 / 0.91 (DeiT-S), 0.90 / 0.96 / 0.93 (DeiT-B).}
\label{tab:rankq}
\footnotesize
\setlength{\tabcolsep}{2.4pt}
\begin{tabular}{llcccccc}
\toprule
 & & \multicolumn{3}{c}{DeiT-S} & \multicolumn{3}{c}{DeiT-B}\\
\cmidrule(lr){3-5}\cmidrule(lr){6-8}
set & budget & add & pair & lin & add & pair & lin\\
\midrule
\multirow{3}{*}{band} & 70\% & 0.66 & 0.80 & 0.73 & 0.74 & 0.83 & 0.77\\
 & 55\% & 0.66 & 0.78 & 0.68 & 0.60 & 0.63 & 0.61\\
 & 40\% & 0.63 & 0.68 & 0.67 & 0.53 & 0.63 & 0.57\\
\midrule
\multirow{7}{*}{near} & 70\% & 0.64 & 0.81 & 0.49 & 0.69 & 0.85 & 0.74\\
 & 65\% & 0.53 & 0.81 & 0.46 & 0.29 & 0.59 & 0.41\\
 & 60\% & 0.46 & 0.53 & 0.53 & 0.49 & 0.67 & 0.53\\
 & 55\% & 0.52 & 0.77 & 0.58 & 0.11 & 0.31 & 0.19\\
 & 50\% & 0.43 & 0.74 & 0.65 & $-0.18$ & 0.13 & $-0.08$\\
 & 45\% & 0.12 & 0.53 & 0.39 & $-0.19$ & 0.19 & $-0.06$\\
 & 40\% & 0.24 & 0.56 & 0.51 & 0.02 & 0.24 & 0.07\\
\bottomrule
\end{tabular}
\end{table}

All three predictors rank worse inside a budget than on the small random structures, and worse near the hybrid's
solution than across the band. The pairwise objective is the best ranker throughout, and lin lies between add and
pair at aggressive budgets. The two models differ sharply near the solution. On DeiT-S every predictor keeps a positive
correlation at every budget ($\tau\ge0.12$ for add, $\ge0.39$ for lin). On DeiT-B at 55\% and below, the correlation
of add and lin with the measured KL is near zero or slightly negative, and that of the pairwise objective is at most
0.31: among structures that differ from the solution in one or two layers, none of the predictors can tell better from
worse. This is the regime in which the search outcome depends on real evaluations (Section~\ref{sec:seeds}) and in
which ranking moves by lin is no better than ranking them at random (Section~\ref{sec:ablation}).

\subsection{What makes the hybrid work? An ablation}
\label{sec:ablation}
The hybrid combines three ingredients: a starting point that may come from lin, a ranking of moves by lin, and real
evaluations. The equal-cost baselines of Section~\ref{sec:setup} share the starting points and differ in the
proposal mechanism, so they do not separate the first two. We therefore ablated them on full ImageNet-1k with everything
else fixed: the neighborhood (all feasible one- and two-layer moves), the 1{,}000-image subset, chunks of 12
measurements, and the number of real structure evaluations, which equals the hybrid's at every budget (74--243). Unlike
the hybrid, a variant does not stop after four unproductive chunks but keeps measuring ranked moves until the
evaluation budget is spent, so all variants receive exactly the same number of evaluations. Two factors change:
\begin{itemize}
\item \emph{Ranker}: lin (as the hybrid), add (the sum of single-block scores), or a uniformly random order (three
  search seeds).
\item \emph{Starts}: the best of \{lin, ILP, greedy\} by real KL (as the hybrid), or of \{ILP, greedy\} only.
\end{itemize}
With lin ranking and all starts, the variant reproduces the hybrid's structures at all fine-tuned budgets. All
variants were fine-tuned at 55\%--40\% with three seeds; for the random ranker, search seed 0 was fine-tuned at every
budget and seeds 1 and 2 at DeiT-B 50\% and 40\%.

\begin{table}[t]
\centering
\caption{Ablation of the hybrid on full ImageNet-1k, after fine-tuning (three fine-tuning seeds). First row: hybrid
(lin ranking, all starts) top-1; other rows: difference to it (points). ``-lin'': without the lin start. Random:
search seed 0. After Holm adjustment over these 56 seed-0 comparisons, every nonzero difference is significant except
those marked $^\circ$. Rows ``3-seed mean'': mean over random search seeds 0--2 at DeiT-B 50\% and 40\% (seed
ranges $-0.65$ to $+0.57$ and $-0.51$ to $+2.93$ with all starts, $-1.49$ to $+0.81$ and $+2.09$ to $+3.50$ without
the lin start); these means are descriptive and not tested.}
\label{tab:ablation}
\footnotesize
\setlength{\tabcolsep}{2.2pt}
\begin{tabular}{lcccc}
\toprule
 & 55\% & 50\% & 45\% & 40\%\\
\midrule
\multicolumn{5}{l}{\emph{DeiT-S}}\\
hybrid & 71.30 & 69.34 & 66.27 & 62.27\\
lin, -lin & $0.00$ & $-0.41$ & $0.00$ & $0.00$\\
add & $+0.63$ & $-0.38$ & $-1.25$ & $-2.73$\\
add, -lin & $-0.69$ & $-0.98$ & $-2.15$ & $-3.20$\\
random & $-0.13^\circ$ & $-0.09^\circ$ & $-0.42$ & $-0.90$\\
random, -lin & $+0.01^\circ$ & $-4.37$ & $-1.07$ & $-4.38$\\
IQP & $-1.06$ & $-1.49$ & $-3.21$ & $-1.81$\\
\midrule
\multicolumn{5}{l}{\emph{DeiT-B}}\\
hybrid & 76.71 & 74.29 & 72.13 & 65.54\\
lin, -lin & $0.00$ & $+0.87$ & $0.00$ & $+2.19$\\
add & $-1.01$ & $-1.85$ & $-3.17$ & $-5.18$\\
add, -lin & $-2.51$ & $-4.03$ & $-10.04$ & $-11.45$\\
random & $-1.21$ & $-0.65$ & $-2.81$ & $+2.93$\\
\quad 3-seed mean & -- & $-0.17$ & -- & $+1.33$\\
random, -lin & $-0.73$ & $+0.81$ & $-1.28$ & $+3.50$\\
\quad 3-seed mean & -- & $-0.58$ & -- & $+2.73$\\
IQP & $-1.16$ & $+0.17^\circ$ & $-1.08$ & $-1.72$\\
\bottomrule
\end{tabular}
\end{table}

Table~\ref{tab:ablation} separates the ingredients.
\begin{itemize}
\item \emph{Effect of the lin starting point.} With lin ranking, dropping the lin start changes the result in three
  of eight settings, by $-0.41$ to $+2.19$ points; at DeiT-B 40\% it improves the result. With add or random ranking it
  costs up to 6.9 and 4.3 points, respectively, so the start matters most when the ranker is weak.
\item \emph{lin ranks better than add.} With the same starts, lin ranking beats add ranking in seven of eight settings,
  by up to 5.2 points; the exception is DeiT-S 55\% ($-0.63$). The sum of single scores, which is what an additive
  pipeline would use, is the weakest ranker throughout.
\item \emph{Against random ranking, the value of lin depends on the model and the budget.} On DeiT-S lin is better by
  0.1--0.9 points with the same starts (not significant at 55\% and 50\%) and by up to 4.4 points without the lin
  start (equal at 55\%). On DeiT-B it is better at 55\%
  and 45\% (by 0.7--2.8 points), not consistently different at 50\% (random seeds range from $-1.49$ to $+0.81$), and
  \emph{worse} at 40\%, where five of six random-ranking runs beat the hybrid, by up to 3.5 points.
\end{itemize}
The DeiT-B 40\% result matches the earlier observations at this budget: lin selects structures whose real KL is far
above its prediction (Fig.~\ref{fig:calib}), most of the additivity error of DeiT-B is functional
(Section~\ref{sec:q2}), and evolution with ten times the evaluations ends 3.4 points ahead of the hybrid
(Table~\ref{tab:evo10}), and near the solution lin's ranking is uncorrelated with the measured KL
(Table~\ref{tab:rankq}). In that regime the predictor misdirects the local search: both the lin start and the lin ranking lower the result,
and a random order of moves does better. Elsewhere, ranking with lin is the most reliable of the three proposal mechanisms, and
in all variants the real evaluations, not the predictor, decide which structure is kept.

\subsection{Discussion and limitations}
\paragraph{When does additivity break, and does it matter?}
Down to 60\% latency on the vision models and 75\% on the language model, the IQP's advantage over the additive
search after fine-tuning is at most 1.6 top-1 points on ImageNet-1k (significant in several settings), and at most
0.8 points of zero-shot accuracy and 0.9 points of perplexity on Qwen2.5-0.5B; whether such differences are acceptable
depends on the application. At more aggressive budgets the additive search falls behind. The run-length interventions
of Section~\ref{sec:runlen} point to a reason on DeiT: an additive objective does not penalize removals in adjacent layers, and
forbidding more than two of them in any three adjacent layers removes most of the gap. How much of the loss remains
after fine-tuning depends on the architecture: on DeiT it persists through a 10{,}000-step schedule and with each
method at the learning rate that is best on the evaluation set, and on the
language model through our 2{,}000-step budget; on Swin-T it shrinks to at most about 1.8 points within 500 steps
(Section~\ref{sec:curve}).

\paragraph{Practical guidance.}
For mild budgets, the additive ILP of LANA- or Puzzle-style pipelines lost at most 1.6 points in our tests. For
aggressive budgets, the cheapest fix is a run-length constraint on removals, which is linear and needs no extra
measurements. Beyond that, a modest number of real evaluations on candidates ranked by a cheap predictor is effective,
provided the predictor is validated per model and budget: where the functional part of the error dominates (DeiT-B at
40\%) or the summed logits overshoot (the language model), black-box search with more real evaluations is the safer
choice.

\paragraph{Limitations.}
We study three vision transformers and one small language model; larger language models, where Puzzle operates, are
not covered, and the ImageNet-100 follow-up studies were not repeated on ImageNet-1k. The run-length interventions and
the ablation cover DeiT at 55\%--40\% only, and at the most extreme budget (DeiT-B, 40\%) black-box search finds better
structures than both the IQP and the hybrid. The pairwise model ignores interactions of three or more blocks and
treats unmeasured pairs as additive. The run-length results do not explain why fine-tuning recovers runs of attention
removal on Swin-T and not on the language model. The metric/functional split is a decomposition of the score, not a
causal one. Fine-tuning is not run to convergence, and only the learning rate was tuned per method, with the best rate
selected on the evaluation set. The language-model predictor uses a bucketed vocabulary, latency is measured on one GPU
type, deterministic searches are run once per budget, and the confidence intervals are conditional on the trained
fine-tuning seeds.

\section{Conclusion}
\label{sec:conclusion}
On DeiT-S, DeiT-B, Swin-T and Qwen2.5-0.5B, an additive blockwise search loses at most 1.6 top-1 points to a
pairwise interaction-aware search at the mildest budgets tested and falls far behind at aggressive budgets. On DeiT a
major failure pattern is the stacking of removals over adjacent layers: capping such runs, with the library unchanged, removes
most of the gap. Whether fine-tuning repairs the damage depends on the architecture: on Swin-T the gap shrinks to at
most about 1.8 points within 500 steps, whereas 3.3--9.9 points remain on DeiT after 10{,}000 steps and the gap persists
on the language model within our budget. Splitting the additivity error into a metric and a functional part indicates
when a cheap logit-sum predictor can stand in for pair measurements. A hybrid search that uses such a predictor only to
rank candidates for real evaluation removes the pair-measurement stage and beats the pairwise IQP in 10 of 14
ImageNet-1k settings after Holm adjustment. Its ranking helps wherever the predictor is reliable; at the most
aggressive DeiT-B budget it is not, and black-box search does better.


\bibliographystyle{IEEEtranN}
\bibliography{refs}

\clearpage
\appendix
\section{Latency, cost and equal-cost baselines on ImageNet-100}
Companion to Sections~\ref{sec:setup} and~\ref{sec:q1}.
\begin{figure}[H]
\centering
\includegraphics{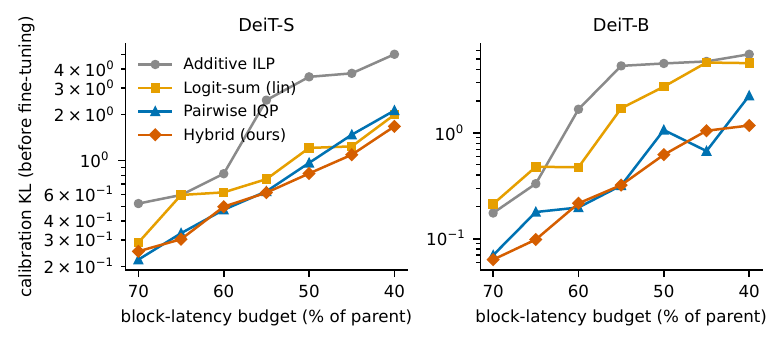}
\caption{Calibration KL of the selected structures before fine-tuning (log scale).}
\label{fig:calib}
\end{figure}
\begin{table}[H]
\centering
\caption{End-to-end latency of the structures selected in the ImageNet-100 experiments, as a fraction of the parent
(RTX 5090, batch 64). Each timing is the median of 100 runs after 20 warm-up runs, as for the block table; the value
shown is the median of three such timings. Parent: 5.84\,ms (DeiT-S), 16.84\,ms (DeiT-B).}
\label{tab:latency}
\footnotesize
\setlength{\tabcolsep}{3pt}
\begin{tabular}{lcccccc}
\toprule
 & \multicolumn{3}{c}{DeiT-S} & \multicolumn{3}{c}{DeiT-B}\\
budget & ILP & IQP & hybrid & ILP & IQP & hybrid\\
\midrule
70\% & 73.5\% & 73.0\% & 73.0\% & 71.1\% & 72.0\% & 71.9\%\\
65\% & 68.2\% & 67.9\% & 68.8\% & 66.6\% & 67.2\% & 66.6\%\\
60\% & 64.1\% & 64.0\% & 64.2\% & 61.8\% & 62.8\% & 62.0\%\\
55\% & 59.0\% & 59.4\% & 59.5\% & 56.9\% & 57.3\% & 57.3\%\\
50\% & 54.4\% & 54.6\% & 54.8\% & 51.6\% & 52.2\% & 52.8\%\\
45\% & 49.5\% & 50.0\% & 49.7\% & 47.6\% & 47.8\% & 48.0\%\\
40\% & 44.8\% & 45.0\% & 44.3\% & 42.5\% & 42.4\% & 42.7\%\\
\bottomrule
\end{tabular}
\end{table}
\begin{table}[H]
\centering
\caption{Search cost beyond the 84 single-block evaluations, in forward images (all seven budgets). IQP: 1{,}389
pairs $\times$ 5{,}000 images. Hybrid: 930 (DeiT-S) or 1{,}843 (DeiT-B) structure evaluations $\times$ 1{,}000
images.}
\label{tab:cost}
\small
\begin{tabular}{lccc}
\toprule
model & IQP pair stage & hybrid & ratio\\
\midrule
DeiT-S & 6.95\,M & 0.93\,M & 13.4\%\\
DeiT-B & 6.95\,M & 1.84\,M & 26.5\%\\
\bottomrule
\end{tabular}
\end{table}
\begin{table}[H]
\centering
\caption{Hybrid versus equal-cost searches on measured KL (hybrid minus random or evolution, top-1 after fine-tuning;
\sig: unadjusted 95\% interval excludes 0). Holm-adjusted wins / ties / losses of the hybrid over the 14 settings:
val100, vs.\ random 2/12/0 and vs.\ evolution 0/14/0; val1k, vs.\ random 10/2/2 and vs.\ evolution 11/2/1.}
\label{tab:baselines}
\footnotesize
\setlength{\tabcolsep}{2.5pt}
\begin{tabular}{lcccc}
\toprule
 & \multicolumn{2}{c}{DeiT-S val100} & \multicolumn{2}{c}{DeiT-B val100}\\
budget & $-$ random & $-$ evolution & $-$ random & $-$ evolution\\
\midrule
70\% & $-0.19$ & $-0.02$ & $+0.67$\sig & $+0.46$\sig\\
65\% & $+0.29$ & $+0.03$ & $+0.29$ & $-0.26$\\
60\% & $-0.05$ & $-0.05$ & $-0.10$ & $-0.11$\\
55\% & $-0.10$ & $-0.10$ & $+0.33$ & $+0.19$\\
50\% & $+1.87$\sig & $+0.63$\sig & $+0.76$\sig & $-0.49$\\
45\% & $-0.21$ & $+0.48$ & $+0.38$ & $-0.74$\sig\\
40\% & $+0.72$\sig & $+0.72$\sig & $+2.12$\sig & $+0.16$\\
\bottomrule
\end{tabular}
\end{table}

\section{Longer fine-tuning at mild budgets (ImageNet-100)}
DeiT-B children of the three mildest budgets, fine-tuned for 10{,}000 instead of 2{,}000 steps (Section~\ref{sec:q1}).
\begin{table}[H]
\centering
\caption{DeiT-B, 10{,}000-step fine-tuning (two seeds), val100 top-1.}
\label{tab:ft10k}
\small
\begin{tabular}{lccc}
\toprule
budget & ILP & IQP & IQP $-$ ILP [95\% CI]\\
\midrule
70\% & 95.66 & 95.99 & $+0.33$ [$+0.01$, $+0.67$]\sig\\
65\% & 95.03 & 95.62 & $+0.59$ [$+0.23$, $+0.95$]\sig\\
60\% & 94.65 & 95.33 & $+0.68$ [$+0.29$, $+1.09$]\sig\\
\bottomrule
\end{tabular}
\end{table}

\section{Hybrid v2: a stronger search lowers KL without consistently improving fine-tuned accuracy}
\label{sec:hybrid2}
Since evolution matched the hybrid on val100, we tried to combine the two: \emph{hybrid~v2} runs regularized
evolution with the same budget of real evaluations; each generation proposes 48 random mutations plus the 8 best lin
moves of the parent and measures only the candidate with the best surrogate score. The surrogate is
$\log\text{lin}(A)$ plus a ridge regression on $\log[\text{act}(A)/\text{lin}(A)]$ fitted online to all measured
structures (both KL values clipped below at $10^{-6}$), with 1{,}113 features: one-hot (layer, option) indicators and
option pairs of layers at distance $\le2$. Hybrid~v2 did find structures with lower pre-fine-tuning KL; on DeiT-B it
is below the hybrid at six of seven budgets, for example 0.22 vs.\ 0.32 at 55\% and 0.68 vs.\ 1.05 at 45\%. After
fine-tuning, however, it shows no consistent gain: on val100 it is significantly better than the hybrid in 1 of 14
settings after Holm adjustment with no significant losses, and on val1k it is significantly worse in 9 of 14 settings, by up to
11.3 points (DeiT-B 45\%). Table~\ref{tab:hybrid2} gives the full comparison with the other searches.

\begin{table}[H]
\centering
\caption{Hybrid v2 versus the other searches: Holm-adjusted counts of significant wins / ties / losses of hybrid v2
over the 14 settings (two models $\times$ seven budgets; ImageNet-100).}
\label{tab:hybrid2}
\footnotesize
\setlength{\tabcolsep}{3pt}
\begin{tabular}{lccccc}
\toprule
 & vs.\ ILP & vs.\ IQP & vs.\ hybrid & vs.\ evolution & vs.\ random\\
\midrule
val100 & 10/4/0 & 6/8/0 & 1/13/0 & 1/13/0 & 3/11/0\\
val1k  & 13/0/1 & 4/4/6 & 1/4/9 & 10/0/4 & 8/0/6\\
\bottomrule
\end{tabular}
\end{table}

\begin{figure}[H]
\centering
\includegraphics{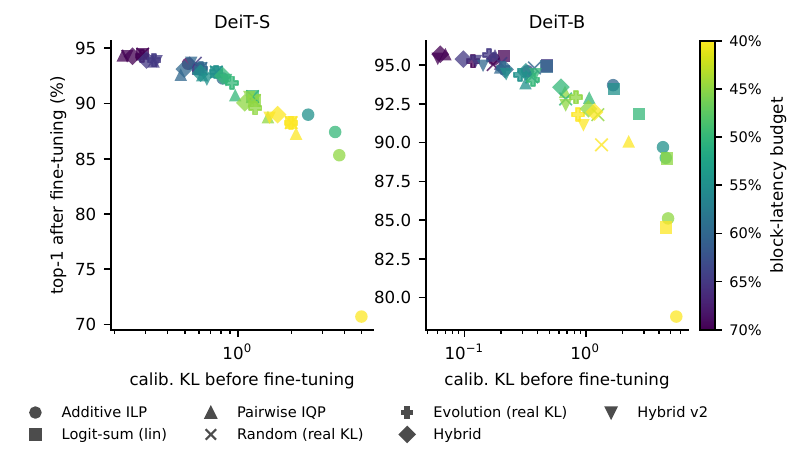}
\caption{Calibration KL before fine-tuning versus val100 top-1 after fine-tuning for all seven searches and budgets
(color). Across budgets, lower calibration KL is generally associated with higher fine-tuned accuracy. Within a
budget, the high-performing structures found by the real-evaluation searches have similar scores, and their order after
fine-tuning is only weakly related to their KL.}
\label{fig:proxy}
\end{figure}

Figure~\ref{fig:proxy} explains why. Across budgets and across clearly different structures, the pre-fine-tuning KL
predicts the fine-tuned accuracy well. This is why the additive ILP is so easy to beat. Among the high-performing
candidate structures found by the six non-additive searches, however, the within-budget Kendall $\tau$ between $-$KL and
fine-tuned top-1 ranges from $-0.57$ to $+1.00$ on DeiT-S, where it is negative at three of seven budgets. On DeiT-B
it ranges from $+0.20$ to $+0.87$. Among these structures, lower pre-fine-tuning KL does not consistently predict higher
post-fine-tuning accuracy. This suggests that, in the present setup, the fidelity of the score rather than the search
limits further gains; the remaining differences between good searches are typically below one point.

\section{Does a fine-tuning-aware score help?}
\label{sec:ftproxy}
A natural fix for the score bottleneck is to score a structure after a \emph{short} fine-tune. We tested this on the
distinct final structures of the seven searches at the seven budgets: 39 on DeiT-S and 49 on DeiT-B (structures
selected by several searches are counted once).
Each structure gets a 500-step global-KD fine-tune with the recipe of Section~\ref{sec:setup} (schedule compressed to
500 steps, identical data order for all structures, two seeds), and calibration KL and top-1 are recorded after 0, 50,
100, 200 and 500 steps on the held-out \emph{calib} images, which the proxy fine-tunes do not see. The cost is 24\,s
(DeiT-S) and 59\,s (DeiT-B) per structure and seed on one RTX~5090.
As before, we report the within-budget Kendall $\tau$ between each score and the 3-seed fine-tuned val100 top-1, over
the six non-additive searches. We also report the effect of using each score to \emph{select} one of the six
structures per budget. The proxy variant to carry forward, calibration top-1 after 200 steps ($\tau=0.80$), was chosen by its
correlation with val100 accuracy on the first DeiT-S proxy seed; the remaining runs, in particular DeiT-B, serve as a
subsequent check. The DeiT-S entries of Table~\ref{tab:ftproxy} average both proxy seeds and therefore include the seed
used for this choice; the DeiT-B entries do not depend on it.

\begin{table}[H]
\centering
\caption{Fine-tuning-aware scores. Mean within-budget Kendall $\tau$ with fine-tuned val100 top-1 over the seven
budgets (minimum in parentheses), and the mean top-1 difference when the structure with the best score among the six
non-additive searches is selected instead of the hybrid's. Two proxy seeds. Rows marked ``50--500 steps'' give the
range of these means over the four checkpoints. The cross-seed agreement row reports the correlation between one
fine-tuning seed and the mean of the other two, an empirical reference for cross-seed agreement.}
\label{tab:ftproxy}
\small
\begin{tabular}{lcccc}
\toprule
 & \multicolumn{2}{c}{DeiT-S} & \multicolumn{2}{c}{DeiT-B}\\
score & $\tau$ & select $-$ hyb & $\tau$ & select $-$ hyb\\
\midrule
calib KL, 0 steps (used so far) & $+0.22$ ($-0.57$) & $-0.10$ & $+0.52$ ($+0.20$) & $+0.17$\\
calib KL, 50--500 steps & $0.00$ to $+0.14$ & $-0.20$ to $-0.14$ & $+0.56$ to $+0.62$ & $+0.15$ to $+0.24$\\
calib top-1, 0 steps & $+0.54$ ($0.00$) & $+0.09$ & $+0.62$ ($+0.33$) & $+0.19$\\
calib top-1, 50--500 steps & $+0.48$ to $+0.70$ & $-0.03$ to $+0.06$ & $+0.50$ to $+0.60$ & $+0.09$ to $+0.19$\\
calib top-1, 200 steps (fixed) & $+0.70$ ($+0.33$) & $+0.06$ & $+0.50$ ($+0.07$) & $+0.09$\\
\midrule
cross-seed agreement & $+0.73$ & --- & $+0.85$ & ---\\
\bottomrule
\end{tabular}
\end{table}

Table~\ref{tab:ftproxy} shows a negative result. Short fine-tuning does not make the KL a better score: on DeiT-S
$\tau$ is near zero after every checkpoint, and on DeiT-B it improves only from 0.52 to about 0.6. The variant
selected on the first DeiT-S seed, top-1 after 200 steps, does not carry over to DeiT-B ($\tau=0.50$, below
calibration top-1 before fine-tuning). The only consistent finding needs no additional forward passes: among the
evaluated structures, calibration \emph{top-1} before fine-tuning ranks better than calibration KL ($\tau$ 0.54
vs.\ 0.22 on DeiT-S, 0.62 vs.\ 0.52 on DeiT-B). Selecting structures by this score among the outputs of the six searches yields mean gains of only 0.09 (DeiT-S) and
0.19 (DeiT-B) percentage points over the hybrid. These gains are small, but we have not assessed them with paired
comparisons, so we do not claim that they are explained by noise. The imperfect cross-seed agreement (0.73 / 0.85)
suggests that fine-tuning variability contributes to the remaining ranking uncertainty, so that at this fine-tuning
budget structures differing by a few tenths of a point are hard to order reliably.

\section{Full ImageNet-1k: Holm-adjusted comparisons of the hybrid}
Companion to Section~\ref{sec:in1k} and Table~\ref{tab:in1k}.
\begin{table}[H]
\centering
\caption{Full ImageNet-1k: wins / ties / losses of the hybrid over each other search (paired tests over the
50{,}000 validation images; seven budgets per model), Holm-adjusted over the 70 comparisons. Last row: unadjusted
totals.}
\label{tab:in1k_wtl}
\footnotesize
\setlength{\tabcolsep}{2.5pt}
\begin{tabular}{lccccc}
\toprule
 & vs.\ ILP & vs.\ IQP & vs.\ lin & vs.\ random & vs.\ evolution\\
\midrule
DeiT-S & 7/0/0 & 6/1/0 & 4/3/0 & 5/2/0 & 4/3/0\\
DeiT-B & 7/0/0 & 4/3/0 & 5/1/1 & 6/1/0 & 5/1/1\\
\midrule
total (14) & 14/0/0 & 10/4/0 & 9/4/1 & 11/3/0 & 9/4/1\\
unadjusted & 14/0/0 & 13/1/0 & 10/3/1 & 11/3/0 & 10/3/1\\
\bottomrule
\end{tabular}
\end{table}

\section{Forbidding whole-block removal}
Companion to Section~\ref{sec:runlen}.
\begin{table}[H]
\centering
\caption{Searches without whole-block removal on DeiT-S (S) and DeiT-B (B) (full ImageNet-1k, 2{,}000-step fine-tune, three seeds). Gap: IQP $-$ ILP. ``Orig.'':
gap with \texttt{block\_skip} allowed (Table~\ref{tab:in1k}). ``Run'': longest run of adjacent layers that
receive the same replacement option, ILP / IQP. \sig: 95\% interval excludes 0.}
\label{tab:noskip}
\footnotesize
\setlength{\tabcolsep}{2.5pt}
\begin{tabular}{llccccc}
\toprule
 & & orig. & \multicolumn{3}{c}{without \texttt{block\_skip}} & \\
\cmidrule(lr){4-6}
model & budget & gap & ILP & IQP & gap [95\% CI] & run\\
\midrule
\multirow{3}{*}{S} & 55\% & $+7.03$ & 62.76 & 69.80 & $+7.04$ [$+6.73$, $+7.35$]\sig & 3 / 2\\
 & 50\% & $+9.66$ & 63.48 & 65.90 & $+2.42$ [$+2.14$, $+2.71$]\sig & 4 / 2\\
 & 45\% & $+6.54$ & 53.47 & 59.30 & $+5.83$ [$+5.55$, $+6.12$]\sig & 9 / 6\\
\midrule
\multirow{4}{*}{B} & 55\% & $+5.79$ & 69.22 & 74.57 & $+5.35$ [$+5.10$, $+5.60$]\sig & 7 / 3\\
 & 50\% & $+8.32$ & 65.06 & 70.11 & $+5.05$ [$+4.78$, $+5.31$]\sig & 7 / 5\\
 & 45\% & $+16.10$ & 61.88 & 66.45 & $+4.58$ [$+4.27$, $+4.87$]\sig & 7 / 6\\
 & 40\% & $+14.62$ & 56.56 & 56.51 & $-0.06$ [$-0.22$, $+0.11$] & 11 / 11\\
\bottomrule
\end{tabular}
\end{table}

\section{Search-seed variance}
Companion to Section~\ref{sec:seeds}.
\begin{table}[H]
\centering
\caption{DeiT-B, full ImageNet-1k, after the 2{,}000-step fine-tune (three fine-tuning seeds): structures of
evolution at $10\times$ (one per search seed) against the hybrid. Seed 0--2: evolution runs. Last column: mean over the three search seeds minus
hybrid, with 95\% paired bootstrap interval (\sig: excludes 0).}
\label{tab:evo10}
\footnotesize
\setlength{\tabcolsep}{2.5pt}
\begin{tabular}{lcccccl}
\toprule
budget & IQP & hyb. & seed 0 & seed 1 & seed 2 & evo.\ $-$ hyb.\ [95\% CI]\\
\midrule
50\% & 74.45 & 74.29 & 74.10 & 75.16 & 74.43 & $+0.27$ [$+0.06$, $+0.48$]\sig\\
45\% & 71.05 & 72.13 & 71.83 & 72.90 & 72.17 & $+0.17$ [$-0.05$, $+0.40$]\\
40\% & 63.83 & 65.54 & 69.40 & 69.66 & 67.79 & $+3.41$ [$+3.12$, $+3.71$]\sig\\
\bottomrule
\end{tabular}
\end{table}
\begin{table}[H]
\centering
\caption{Full ImageNet-1k, top-1 (\%) before fine-tuning of the structure selected by each search. Random and
evolution: mean $\pm$ standard deviation over three search seeds, with the hybrid's number of real evaluations
($1\times$) or ten times as many ($10\times$). IQP and hybrid are deterministic.}
\label{tab:seeds}
\small
\setlength{\tabcolsep}{3.5pt}
\begin{tabular}{llcccccc}
\toprule
model & budget & IQP & hybrid & random $1\times$ & evolution $1\times$ & random $10\times$ & evolution $10\times$\\
\midrule
\multirow{7}{*}{DeiT-S}
 & 70\% & 73.40 & \textbf{73.73} & $71.20\pm0.00$ & $71.20\pm0.00$ & $71.20\pm0.00$ & $72.74\pm0.56$\\
 & 65\% & 70.92 & \textbf{71.57} & $67.69\pm0.00$ & $68.96\pm1.20$ & $67.70\pm0.20$ & $69.83\pm0.31$\\
 & 60\% & 66.23 & \textbf{69.10} & $67.55\pm0.00$ & $67.55\pm0.00$ & $67.55\pm0.00$ & $68.67\pm0.80$\\
 & 55\% & 62.45 & \textbf{66.56} & $64.89\pm0.00$ & $64.89\pm0.00$ & $64.89\pm0.00$ & $64.81\pm0.77$\\
 & 50\% & 58.15 & \textbf{62.35} & $61.49\pm0.00$ & $61.49\pm0.00$ & $61.49\pm0.00$ & $60.24\pm1.82$\\
 & 45\% & 47.67 & \textbf{56.26} & $52.83\pm0.00$ & $52.83\pm0.00$ & $52.83\pm0.00$ & $53.36\pm0.71$\\
 & 40\% & 43.94 & \textbf{46.33} & $36.66\pm0.89$ & $38.61\pm3.12$ & $33.56\pm2.45$ & $43.72\pm2.46$\\
\midrule
\multirow{7}{*}{DeiT-B}
 & 70\% & 78.84 & \textbf{79.05} & $74.47\pm0.55$ & $76.38\pm0.97$ & $76.02\pm0.87$ & $78.33\pm0.50$\\
 & 65\% & 74.96 & \textbf{77.20} & $76.93\pm0.00$ & $76.93\pm0.00$ & $76.93\pm0.00$ & $77.08\pm0.20$\\
 & 60\% & 72.30 & 75.32 & $70.37\pm1.66$ & $73.64\pm0.27$ & $71.26\pm1.04$ & \textbf{75.50}$\pm0.68$\\
 & 55\% & 58.24 & 71.13 & $68.04\pm2.32$ & $69.06\pm1.40$ & $68.47\pm0.76$ & \textbf{71.97}$\pm0.46$\\
 & 50\% & 51.35 & 63.84 & $58.54\pm2.32$ & $65.92\pm1.55$ & $62.57\pm2.76$ & \textbf{68.64}$\pm0.78$\\
 & 45\% & 40.54 & 43.86 & $53.05\pm3.78$ & $58.90\pm1.06$ & $57.55\pm3.04$ & \textbf{64.06}$\pm1.23$\\
 & 40\% & 11.12 & 23.36 & $40.31\pm3.79$ & $46.99\pm2.20$ & $42.54\pm2.46$ & \textbf{56.70}$\pm1.89$\\
\bottomrule
\end{tabular}
\end{table}

\end{document}